\documentclass{qjmam}
\usepackage{graphicx,color,soul,comment}
\usepackage{amsmath,amssymb,enumerate}

\startpage{1}
\DeclareMathAlphabet\mathbit
    \encodingdefault\rmdefault\bfdefault\itdefault
\DeclareOldFontCommand{\bi}{\normalfont\bfseries\itshape}{\mathbit}

\newcommand{\ee}{\end{equation}}

\def\fakebold#1{\relax\ifvmode\leavevmode\fi%
\ifmmode%
\setbox0=\hbox{$#1$}%
\else%
\setbox0=\hbox{#1}%
\fi%
\kern-.02em\copy0 \kern-\wd0%
\kern .04em\copy0 \kern-\wd0%
\kern-.0125em\raise.02em\box0%
}%

\renewcommand{\geq}{\geqslant}
\renewcommand{\leq}{\leqslant}
\renewcommand{\Re}{{\rm Re}}

\newcommand{\R}{\mathbb R}

\newcommand{\kmax}{\kappa^{\max}}

\newcommand{\matlab}{{\sc matlab}}
\newcommand{\granso}{{\sc granso}}

\newcommand{\kappaopt}{\kappa^{\opt}}
\newcommand{\alphaopt}{\alpha_{1}^{\opt}}
\newcommand{\betaopt}{\beta_{1}^{\opt}}

\newcommand{\begineq}[1]{\begin{equation}\label{#1}} % had to change the names, \be is taken
\newcommand{\ba}[1]{\begin{eqnarray}\label{#1}}
\newcommand{\ea}{\end{eqnarray}}
\newcommand{\rf}[1]{(\ref{#1})}
\newcommand{\nn}{\nonumber}

\newcommand{\M}{{\bf M}}
\newcommand{\K}{{\bf K}}
\newcommand{\kappacrit}{\kappa_{{\rm crit}}}
\newcommand{\kappacritalphaonebetaone}{\kappacrit^{\alpha_{1},\beta_{1}}}
\newcommand{\kappacritalphabeta}{\kappacrit^{\alpha,\beta}}
\newcommand{\opt}{*}

\begin{document}

\title[{The strongest column with a follower load and relocatable masses}] {Finding the strongest stable massless column
  with a follower load and relocatable concentrated masses}

\author[Kirillov  and  Overton]{Oleg N. Kirillov  \and  Michael L. Overton}

\address{Northumbria University, Newcastle upon Tyne, NE1 8ST, UK.\\ Email: oleg.kirillov@northumbria.ac.uk}

\extraaddress{Courant Institute of Mathematical Sciences, New York University,\\ New York, NY, 10012, USA. Email: mo1@nyu.edu}

\received{August 2020 (original submission), February 2021 (revised submission)}

\maketitle

\eqnobysec

%\section{Introduction}
%This document shows how you should set out the preamble of your paper if
%using the QJMAM.cls file. Please set the rest of your paper using your
%normal method. If a maketitle error occurs this can be ignored.
%
%
%%%% \smartqed  % flush right qed marks, e.g. at end of proof
%

%\rule[1mm]{2em}{.15mm}
%

\begin{abstract}
 We consider the problem of optimal placement of concentrated masses along a massless elastic column that is clamped at one end and loaded by a nonconservative follower force at the free end. The goal is to find the largest possible interval such that the variation in the loading parameter within this interval preserves stability of the structure. The stability constraint is nonconvex and nonsmooth, making the optimization problem quite challenging.
We give a detailed analytical treatment for the case of two masses, arguing that the optimal
parameter configuration approaches the flutter and divergence boundaries of the stability region simultaneously.
Furthermore, we conjecture that this property holds for any number of masses, which in turn suggests
a simple formula for the maximal load interval for $n$ masses. This conjecture is strongly supported by extensive computational results, obtained
using the recently developed open-source software package \granso\ (GRadient-based Algorithm for Non-Smooth Optimization)
to maximize the load interval subject to an appropriate formulation of the nonsmooth stability constraint.
We hope that our work will provide a foundation for new approaches to classical long-standing problems of stability optimization
for nonconservative elastic systems arising in civil and mechanical engineering.
\end{abstract}

%\keywords{Stability \and Optimization \and GRANSO \and Beck column \and Pfl\"uger column \and Follower force \and Multiple eigenvalue \and Singular points}
%\mo{What do you think of Quarterly Journal of Mechanics and Applied Math as a possible journal? Isn't this quite well known? It's published
%by Oxford which is nice. Do you know any of the editors? SIAM J.\ Applied Math.\ and IMA J.\ Applied Math.\ could be other possibilities,
%but although I know some of the editors, none of them are anywhere near the field of mechanics. }

% \bigskip

%\mo{What about this as a possible title? Pfl\"uger is hard to say and anyway it seems Dzhanelidze is more accurate, and that's even worse.
%Also, it seemed important to introduce the word ``stable''.}

%\mo{My changes are in red. If you like you can use the macro} \verb@\oleg{...}@ \oleg{to make your changes in blue}.

\bigskip


\section{Introduction}\label{sec:intro}

Consider an elastic Euler-Bernoulli beam clamped at one end and loaded at the tip by a follower force \cite{B1952,CM1979}. The
follower force is defined as a force with the line of action that always coincides with the tangent line to the neutral axis of the deformed beam at its free end, much like a rocket thrust \cite{SLK2019}. The follower force does not depend on the velocity of the beam. However, it cannot be derived from a potential: the work done by the follower force along a closed contour is non-zero \cite{Ziegler53a,Ziegler53b}. This structure is frequently called the Beck column \cite{B1952,CM1979}. A straight form of the Beck column is in a stable equilibrium when the follower force is absent or relatively small. Nevertheless, at some sufficiently large value the follower force excites exponentially growing oscillations of the beam that are known as flutter instability \cite{B1963,K2013}.

Flutter is critically important both for safety of engineering structures interacting with fluid flows and for efficiency of energy harvesting devices that are based on the fluid-structure interactions. Recent years have seen an increasing interest in the Beck column in the modelling of biological filaments and their artificial biomimetic  analogues, i.e., hair-like slender microscale structures that play an important part
in such biological processes as swimming, pumping, mixing, and cytoplasmic streaming by performing rhythmic, wave-like motion that usually sets in via flutter instability  \cite{BD2016,G2017,FGG2021,ZS2019b,ZS2019a}.

Structural optimization of the Beck column against instabilities, including flutter and buckling (or divergence instability), is usually formulated as a problem on a redistribution of the material of the column of a given density under an isoperimetric constraint fixing the volume of the column in order to maximize the range of variation of the follower load corresponding to the stable structure. In the literature many specific numerically optimized shapes of the Beck column have been reported \cite{GZ1988,C1975,HW1980,KK1980,BF2018,MB1981} with the maximal critical dimensionless load reaching the values of $p\approx 100.00$ \cite{BF2018}, $p\approx 139.30$ \cite{LS2000cs}, $p\approx 143.59$ \cite{R1994}, and $p\approx 148.62$ \cite{TF2007}, which significantly improve upon the critical load $p\approx 20.05$ of the uniform column with a constant cross-section (see Appendix \ref{PflugerApp} for the definition of $p$). Nevertheless, none of these designs is proven to be a global or even a local optimizer. Such a proof would be difficult to obtain because
the problem of structural optimization of the critical flutter load for the elastic Beck column is both nonconvex and nonsmooth \cite{KS2002pmm,KS2002vmgu}.

Indeed, the elastic Beck column is a time-reversible dynamical system in which the transition from stability to flutter instability generically happens via the reversible-Hopf bifurcation, i.e., through the formation of a double imaginary eigenvalue with a Jordan block at the stability boundary and its subsequent splitting when parameters enter the instability region \cite{RMN1996}. Codimension-1 parts of the stability boundary are thus smooth hypersurfaces corresponding to double imaginary eigenvalues with a Jordan block (provided that the remaining eigenvalues are simple and imaginary) \cite{K2013,KS2001,KS2004}. These hypersurfaces can meet each other at sets of higher codimension such as intersections, cuspidal edges and points, conical points etc.; see \cite{K2013} for a full classification of generic singularities on the stability boundary of mechanical systems with non-potential positional forces.  The unavoidable singularities linked to multiple eigenvalues is the main reason for nonsmoothness of the merit functionals in the optimization of such systems, including the Beck column, with respect to stability criteria \cite{KO2013,K2011ap}.

Many studies report on the phenomenon of overlapping of eigenvalue curves that accompanies the process of optimization of the Beck column. The eigenfrequencies plotted as functions of the load exhibit sudden crossings during the optimization that lead to transfer of instability between modes and to a discontinuous change in the merit functional \cite{KS1998,C1975,HW1980,KK1980,BF2018,MB1981,R1994,TF2007,LS1999,KT2007,KZ1963}. The high sensitivity of the optimized design to variation of parameters is caused by the nonconvexity of the stability domain \cite{KS2002pmm,KS2002vmgu}. For this reason the unambiguous determination of the optimal design of the Beck column by numerical procedures typically used in civil and mechanical engineering remains a challenge \cite{BF2018,R1994,TF2007}.

All of the phenomena described above were also observed in simplified settings with the uniform Beck column carrying relocatable lumped masses \cite{DL1955,KZ1963,L1997,H1967,LL1970,KL1974a,KL1974b,K1977}. Nevertheless, to the best of our knowledge, no rigorously proven local or global optimal solutions or credible numerical guesses exist in the literature even in the problems of optimal localization of point masses along elastic beams loaded by the follower force.

Structures loaded by follower forces have long been questioned for their practical realization \cite{SLR1999,SLR2002}, despite an evident  example given by flexible missiles \cite{S1975,PM1985,KS1998}. In the 1970-90s, Sugiyama et al.\ used solid rocket motors to demonstrate flutter of cantilevers under a follower thrust on relatively short (several seconds) time intervals  \cite{SLK2019,S1995,Langthjem2012}. A mechanism recently invented by Bigoni and Noselli produces a frictional follower force \cite{BN2011,CDB2020} and enables experimental realization of fluttering cantilevered rods under follower loads on virtually infinite time intervals \cite{BK2018,B2018}.
These practical realizations differ from the classical Beck column, however, by the presence of a finite-size loading unit at the tip of the cantilever and therefore are better described by the model of the Pfl\"uger column \cite{P1955,DL1955}, which is the Beck column with a point mass at the loaded end; see the left panel of Fig.~\ref{fig1} in  Appendix \ref{PflugerApp}.

In recent mechanical laboratory experiments with follower forces \cite{BK2018,B2018}, the ratio of the
end mass to the mass of the column was chosen to be very large, approaching the so-called Dzhanelidze limit corresponding to a massless column \cite{B1963}. The instability thresholds obtained in these experiments were in a very good agreement with the theoretical predictions based on the Pfl\"uger model.
In the Dzhanelidze limit, the mathematical model is reduced to a system of ordinary differential equations \cite{BS2010,KR2001,I2013,KR2014}. The works \cite{B1963,DL1955,KZ1963} considered stability of a massless Pfl\"uger column with an additional relocatable mass.
A recent work \cite{KR2014} corrected some of the results reported in \cite{B1963} and proposed extending the model to incorporate several relocatable masses.

The primary purpose of our paper is to study this last variant, the Pfl\"uger model in the Dzhanelidze
massless limit with relocatable point masses, in detail. One reason is that this comparatively simple but still mechanically meaningful
model allows a detailed analytical treatment of the case of two masses, providing a benchmark for
numerical optimization carried out for $n$ masses.  A second advantage of studying
the discrete mass model instead of the classical Beck column is that it does not require Galerkin or finite element discretization, and hence the number of optimization variables is small (only $2n-1$).
Nonetheless, the problem of maximizing the load interval subject to the stability constraint is far from trivial
because of the nonconvexity and nonsmoothness  (in fact, non-Lipschitzness) of the constraint, so even this simplified model provides a good test
of how much insight we can obtain using nonsmooth optimization techniques.
Our first contribution, presented in Section \ref{sec:analytical},
is to give a detailed analytical treatment for the case of two masses, arguing that the optimal
parameter configuration approaches the flutter and divergence boundaries simultaneously.
Furthermore, we conjecture that this property holds for any number of masses, which in turn suggests a simple formula for the
optimal load interval for $n$ masses. Our second contribution, in Section \ref{sec:experimental}, is to present a practical numerical
formulation of the stability constraint and to maximize the load interval subject to this constraint using modern techniques for nonsmooth, nonconvex optimization,
employing a recently developed open-source software package, \granso\ (GRadient-based Algorithm for Non-Smooth Optimization) \cite{CMO2017,Mit2020}. As well as verifying our analytical solution for two masses, these computations strongly support the formula for the conjectured optimal load interval for $n$ masses.
We hope that our techniques and results will provide a foundation and inspiration for new approaches to classical long-standing problems of
stability optimization for nonconservative elastic systems arising in civil and mechanical engineering.

\section{A massless elastic column with $n$ concentrated masses}\label{sec:weightless}

It is convenient to first consider the simple model of the Pfl\"uger column without relocatable masses, with
zero mass per unit length and zero point mass at the free end of the column (see Appendix  \ref{PflugerApp} for details).
Then, the boundary value problem \rf{dlbp}, \rf{dlbc} takes the form
\begineq{dlbz}
\partial_{\xi}^4 f +\kappa^2 \partial_{\xi}^2 f=0,
\end{equation}
\begineq{dlbz1}
f(0)=0,\quad \partial_{\xi}f(0)=0,\quad
\partial_{\xi}^2 f(1)=0,\quad \partial_{\xi}^3 f(1)=0,
\end{equation}
where
\begineq{dck}
\kappa^2=p,
\end{equation}
with $p$ given in \eqref{dlq}.

Following \cite{B1963,DL1955,KR2014}, consider the case when a concentrated constant force $F$ is acting in a direction perpendicular to the non-deformed column at the point $s=\alpha l$.  Introducing the dimensionless version of the force parameter, $\phi=\frac{F l^2}{EI}$,
we seek the
general solution to the equation \rf{dlbz} in the form \cite{B1963,DL1955,KR2014}
\ba{gsa}
f(\xi)=u(\xi)+\left\{\begin{array}{r}
         0, \quad \xi\in[0,\alpha) \\
         v(\xi), \quad \xi\in[\alpha,1]
       \end{array}\right.
\ea
where
$$
u(\xi)=A\sin{\kappa\xi}+B\cos{\kappa\xi}+C\xi+D
$$
and
$$
v(\xi)=A_1\sin{\kappa \xi}+B_1\cos{\kappa \xi}+C_1\xi+D_1.
$$
To determine the coefficients $A_1$, $B_1$, $C_1$, and $D_1$, we require that
\ba{glu}
&u(\alpha)=f(\alpha),\quad \partial_{\xi}u(\alpha)=\partial_{\xi}f(\alpha),&\nn\\
&\partial^2_{\xi}u(\alpha)=\partial^2_{\xi}f(\alpha),\quad
\partial^3_{\xi}f(\alpha)-\partial^3_{\xi}u(\alpha)=\phi.&
\ea
This yields
\begineq{fv}
v(\xi) = \frac{(\xi-\alpha)\kappa-\sin((\xi-\alpha)\kappa)}{\kappa^3}\phi.
\end{equation}
Taking \rf{fv} into account in the general solution \rf{gsa} and then substituting $f(\xi)$ into the
boundary conditions \rf{dlbz1}, we find the coefficients $A$, $B$, $C$, and $D$ to obtain
\begineq{fu}
u(\xi) = \frac{\sin(\kappa \alpha)-\xi\kappa\cos(\kappa \alpha)+\sin((\xi-\alpha)\kappa)}{\kappa^3}\phi.
\end{equation}

    \begin{figure}[t]
    \begin{center}
    \includegraphics[angle=0, width=0.55\textwidth]{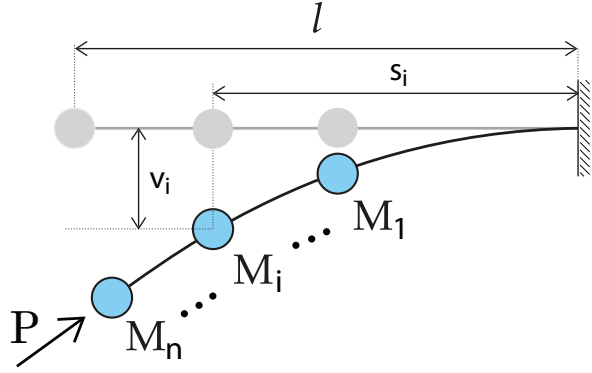}
    \end{center}
    \caption{The massless Beck column loaded by the follower force $P$ with $n$ concentrated masses $M_1$,
    $\ldots$, $M_i$, $\ldots$, $M_n$ attached \cite{K1977,KR2014}. }
    \label{fig2}
    \end{figure}

Let us now assume that the massless cantilevered column loaded by the follower force at its free end carries $n$ concentrated masses
with the mass $M_n>0$ fixed at the loaded end; see Fig.~\ref{fig2}.
The masses $M_i\ge 0$, $i=1,\ldots,n-1,$ are located at the distances $s_i<l$ from the clamped end of the column.
Let $v_i$ be a transversal displacement of the mass $M_i$ from the equilibrium configuration, as shown in Fig.~\ref{fig2}.
Introducing the dimensionless displacements of the masses, $w_i$, the distances, $\alpha_i$, and the mass ratios, $\mu_i$, as
\begineq{dlp}
w_i=\frac{v_i}{l}, \quad \alpha_i=\frac{s_i}{l},\quad \mu_i=\frac{M_i}{M_n}, \quad i=1,\ldots,n,
\end{equation}
we write the equations of motion of the masses \cite{B1963,DL1955,KR2014}
\ba{emm}
w_1&=&-\gamma_{11} \mu_1 \frac{d^2 w_1}{d \tau^2}-\gamma_{12} \mu_2 \frac{d^2 w_2}{d \tau^2}-\ldots -\gamma_{1n} \mu_n \frac{d^2 w_n}{d \tau^2},\nn\\
w_2&=&-\gamma_{21} \mu_1 \frac{d^2 w_1}{d \tau^2}-\gamma_{22} \mu_2 \frac{d^2 w_2}{d \tau^2}-\ldots -\gamma_{2n} \mu_n \frac{d^2 w_n}{d \tau^2},\nn\\
 &\vdots & \nn\\
w_i&=&-\gamma_{i1} \mu_1 \frac{d^2 w_1}{d \tau^2}-\gamma_{i2} \mu_2 \frac{d^2 w_2}{d \tau^2}-\ldots -\gamma_{in} \mu_n \frac{d^2 w_n}{d \tau^2},\nn\\
 &\vdots & \nn\\
w_n&=&-\gamma_{n1} \mu_1 \frac{d^2 w_1}{d \tau^2}-\gamma_{n2} \mu_2 \frac{d^2 w_2}{d \tau^2}-\ldots -\gamma_{nn} \mu_n \frac{d^2 w_n}{d \tau^2},
\ea
where the dimensionless time $\tau$ is defined now as
\begineq{dt2}
\tau=t\sqrt{\frac{EI}{M_n l^3}}.
\end{equation}
Note that $\alpha_{1}\leq \alpha_{2}\leq \ldots \leq \alpha_{n}=1$ and $\mu_{n}=1.$
The coefficient $\gamma_{ij}$ is the displacement of the mass $\mu_i$ as a result of application to the column of a unit force $\phi=1$ at the point $\alpha_j$. According to \rf{gsa} with the functions \rf{fv} and \rf{fu} the coefficient $\gamma_{ij}$  is given by
$\delta_{ij}/\kappa^3$, where
\ba{deltaij}
\delta_{ij}&=&{\sin(\kappa \alpha_j)-\alpha_i\kappa\cos(\kappa \alpha_j)+\sin((\alpha_i-\alpha_j)\kappa)}\nn\\
& & + \left\{\begin{array}{r}
         0, \quad i \leq j  \\ % \alpha_i\in[0,\alpha_j) \\
         (\alpha_i-\alpha_j)\kappa-\sin((\alpha_i-\alpha_j)\kappa), \quad i>j. \\ % \alpha_i\in[\alpha_j,1]
       \end{array}\right.
\ea
Separating time with the ansatz $w_i=u_ie^{\sigma \kappa^{3/2}\tau}$ we arrive at the
eigenvalue problem
\begineq{eip}
({\M}\sigma^2+{\K})u=0,
\end{equation}
where $u=(u_1,u_2,\ldots, u_n)$, {$\K$} is the $n\times n$ unit matrix, and
\begineq{mass}
{\M}=\left(
  \begin{array}{llll}
   \mu_1 \delta_{11}  &  \mu_2\delta_{12}  & \cdots & \mu_n\delta_{1n}\\
   \mu_1 \delta_{21}  &  \mu_2\delta_{22}  & \cdots & \mu_n\delta_{2n}\\
   \vdots & \vdots & \ddots & \vdots \\
   \mu_1 \delta_{n1}  &  \mu_2\delta_{n2}  & \cdots & \mu_n\delta_{nn}\\
  \end{array}
\right) ,
\end{equation}
where, as already noted, $\mu_{n}=1$.
The eigenvalues $\sigma_{k}$ are given by
\begineq{sigma-lambda}
             \sigma_{k} = \pm \sqrt{-\lambda_{k}^{{-1}}}
\end{equation}
where the $\lambda_{k}$ are the eigenvalues of the matrix $\M$.

The trivial equilibrium of the circulatory system \rf{emm} is stable if and only if the eigenvalues $\sigma_{k}$ are imaginary and semisimple
(i.e., the algebraic and geometric multiplicity are equal),
or equivalently, the $\lambda_{k}$ are real, positive and semisimple.
Cases with a multiple imaginary eigenvalue $\sigma_{k}$ with a Jordan block
(i.e., with the algebraic multiplicity exceeding the geometric multiplicity) lie
on the boundary between the stability and flutter domains. In the generic case the crossing of this stability boundary is accompanied by merging of two simple imaginary eigenvalues into a double imaginary eigenvalue with a Jordan block, indicating the onset of the reversible-Hopf bifurcation or flutter \cite{B1963,K2013,KS2001,KS2004}. Non-oscillatory instability or divergence corresponds to one or more
positive real eigenvalues $\sigma_{k}$ and in this model it generically sets in when two conjugate simple imaginary eigenvalues meet at infinity, split and turn back towards the origin along the real axis in the complex plane \cite{O1972,SKK1976,KR2014}.

Summarizing, for a given number of masses $n$, the eigenvalue problem  \rf{eip} is defined by \rf{deltaij} and \rf{mass}, which
depend on the given load $\kappa$ and the parameters  $\alpha_{i}$ and $\mu_{i}$, $i=1,\ldots,n-1$, defined in \rf{dlp} (as $\alpha_{n}=\mu_{n}=1$).
It is convenient to use the parameterization
\begineq{tanbeta}
          \mu_{i}=\tan\beta_{i},  \quad \beta_{i}\in[0,\pi/2), \quad i=1,\ldots,n-1.
\end{equation}
Given $\alpha_{i},\beta_{i}$, $i=1,\ldots,n-1$, let us define $\kappacritalphabeta$  as the largest value such that the
eigenvalues $\sigma_{k}$ (which depend on $\alpha_{i}$, $\beta_{i}$ and $\kappa$)
are imaginary for all $\kappa\in[0,\kappacritalphabeta]$.
Our goal is to find the supremum of  $\kappacritalphabeta$ over all
parameters $\alpha_{i} \in [0,1]$ and $\beta_{i}\in[0,\pi/2)$, $i=1,\ldots,{n-1}$.
We begin with the case $n=2$, where we propose an analytical solution.

\section{Analytical derivation of the supremal load interval for the massless  column carrying two
concentrated masses}\label{sec:analytical}

When $n=2$, the massless
 column carries a relocatable mass $M_{1}$ between the clamped end and the free end of the rod
with mass $M_2$ fixed at the free end.
There are two parameters, $\alpha_{1}$ and $\beta_{1}$. Expression \rf{deltaij}
allows us to find the coefficients $\delta_{ij}$ in the explicit form, cf. \cite{B1963,KR2014},
\ba{entries2}
\delta_{11}&=&\sin(\kappa \alpha_1)-\kappa \alpha_1 \cos(\kappa \alpha_1)\nn\\
\delta_{12}&=&\sin(\kappa)-\kappa \alpha_1\cos(\kappa)-\sin(\kappa(1-\alpha_1))\nn\\
\delta_{21}&=&\sin(\kappa\alpha_1)-\kappa\cos(\kappa \alpha_1)+\kappa(1-\alpha_1)\nn\\
\delta_{22}&=&\sin(\kappa)-\kappa\cos(\kappa).
\ea

    \begin{figure}
    \begin{center}\
    \includegraphics[angle=0, width=0.47\textwidth]{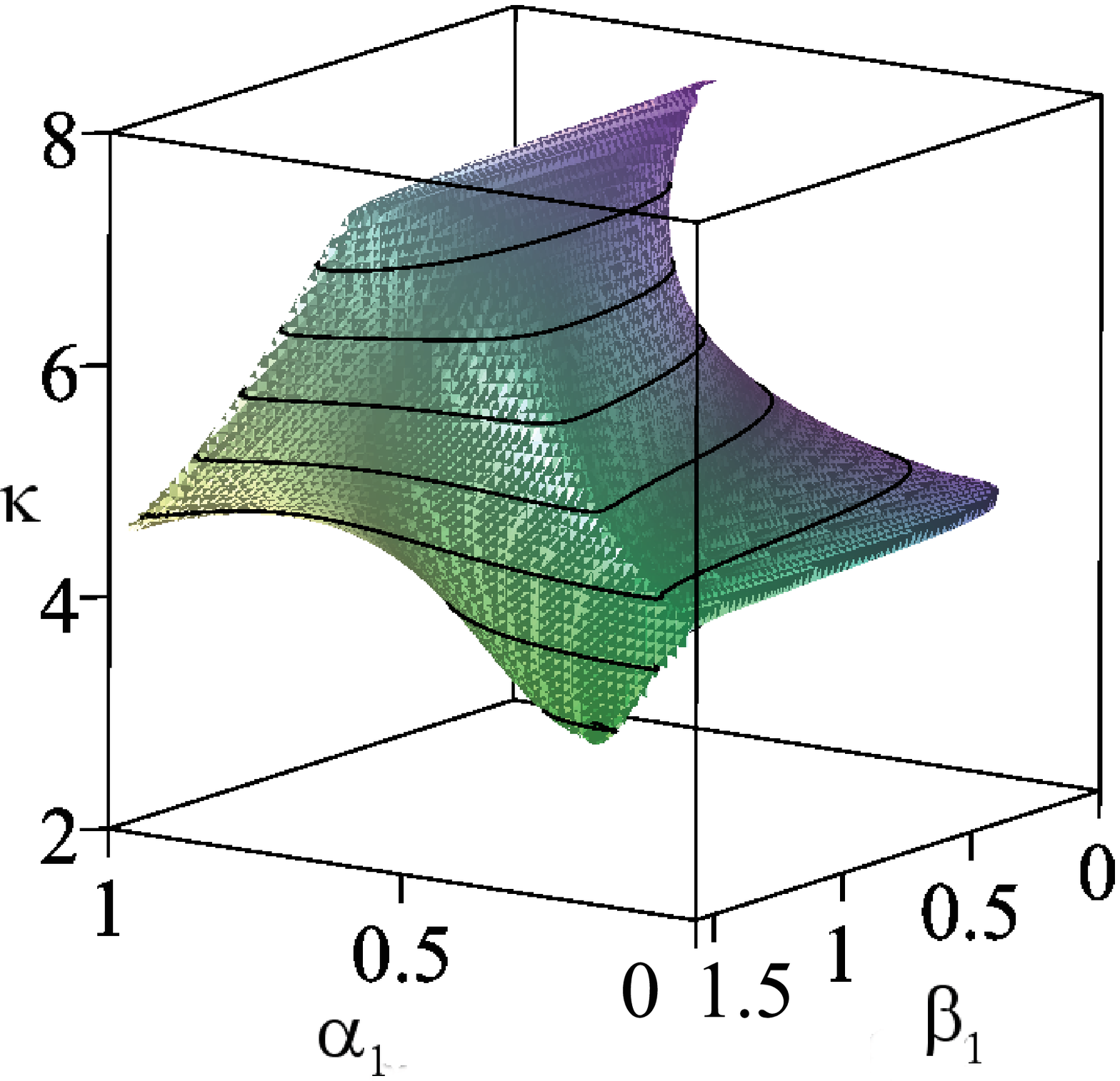}
    \includegraphics[angle=0, width=0.47\textwidth]{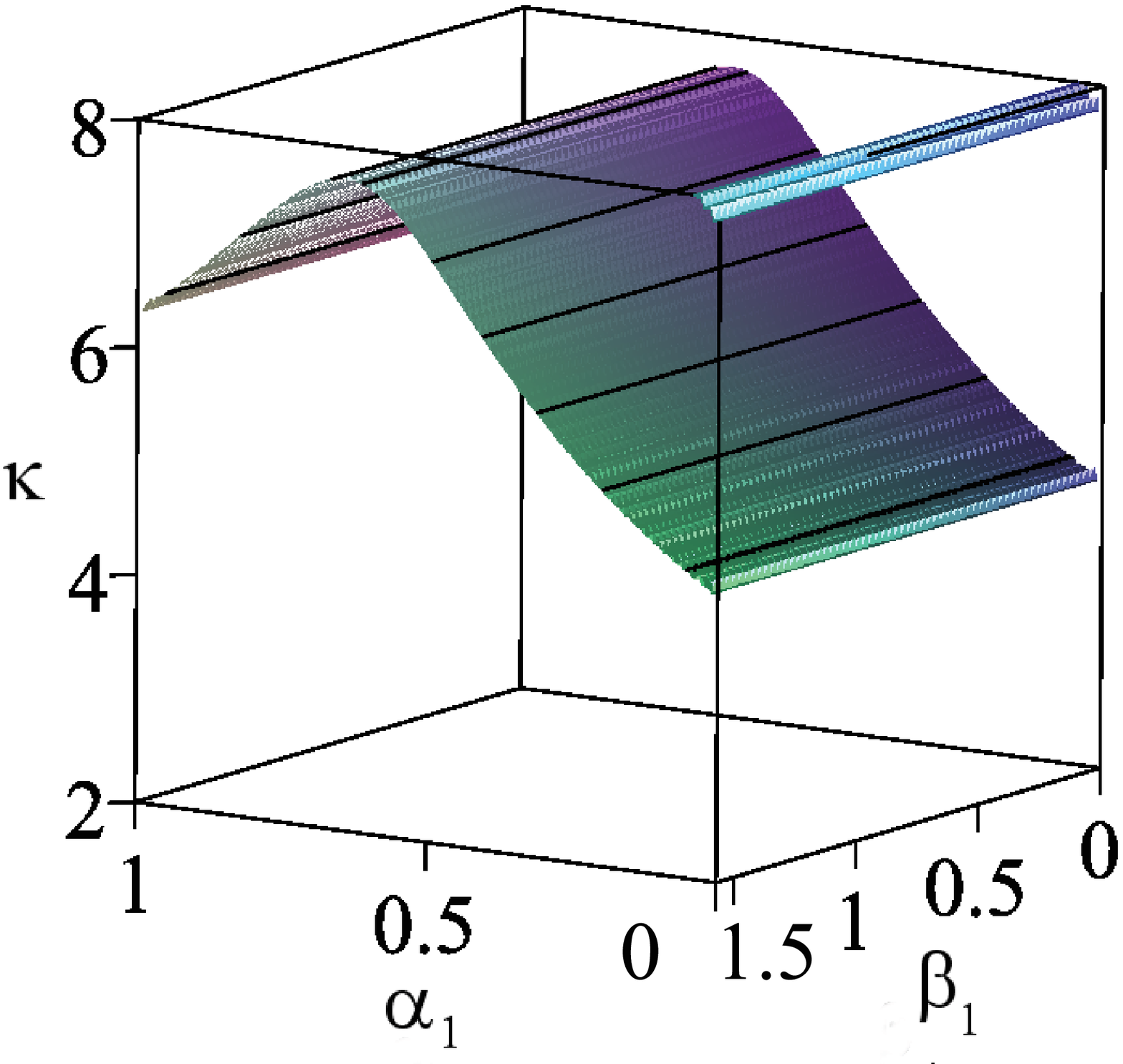}
    \end{center}
    \caption{
    The case of $n=2$ concentrated masses. (Left) the flutter domain is a finite solid set in the
    $(\alpha_1,\beta_1,\kappa)$ space, enclosed within the singular surface defined by \rf{discra}.
    (Right) The divergence domain lies above the  boundary set defined by \rf{divb}.
    For a given $(\alpha_1$, $\beta_1)$, the critical value of the load parameter, $\kappacritalphaonebetaone$, is the minimal value of
   $\kappa$ that satisfies either \rf{discra} or \rf{divb}, as this is the length of the longest vertical line segment rising from the point
   $(\alpha_{1},\beta_{1},0)$ that does not enter either the flutter or divergence domain.
    Consequently, this is the largest value $\tilde\kappa$ such that the
   column is stable for all $\kappa \in [0,\tilde\kappa)$.
   The optimization problem to be solved is to find the supremum
   of $\kappacritalphaonebetaone$ over all $\alpha_{1}\in[0,1], \beta_{1}\in [0,\pi/2)$.}
    \label{fig3}
    \end{figure}

As we will see, already in this simplest possible mechanical system, the subdivision of the parameter space into the domains of stability, flutter instability, and divergence instability is highly nontrivial. However, we will be able to explore it completely and find an apparent
supremum of the critical load parameter defining the longest stability interval $[0,\kappacritalphaonebetaone]$
in the space of parameters $\alpha_{1}\in[0,1]$, $\beta_{1}\in[0,\pi/2)$.

In general, the stability map for a mechanical system with the characteristic polynomial $p(\sigma)=\det(\M\sigma^2+K)$ can be obtained with the use
of the Gallina criterion \cite{K2013,G2003,B2011} that is based on the investigation of the discriminant of the polynomial. For $n=2$, $p(\sigma)$ is a biquadratic function
\ba{charp}
p(\sigma)&=& \sigma^4\tan\beta_1\left\{\kappa(\alpha_1-1)(\sin\kappa-\kappa\alpha_1\cos\kappa+\sin(\kappa\alpha_1-\kappa))\right.\nn\\
&&\qquad -\left.\sin(\kappa(\alpha_1-1))(\sin(\kappa\alpha_1)-\kappa\cos(\kappa\alpha_1)-\kappa(\alpha_1-1))\right\}\nn\\
&&+ ~ \sigma^2\left[\tan\beta_1\left(\sin(\kappa\alpha_1)-\kappa\alpha_1\cos(\kappa\alpha_1)\right)-\kappa\cos\kappa+\sin\kappa\right]+1.
\ea

Notice that the coefficient at the leading power of $\sigma$ in the polynomial \rf{charp} is nothing else but $\det \M$; see \cite{K2013}.
The system loses stability by divergence as soon as $\det \M=0$, which yields the following equation determining the
divergence boundary:
\begineq{divb}
\frac{\sin\kappa-\kappa\alpha_1\cos\kappa+\sin(\kappa\alpha_1-\kappa)}{\sin(\kappa\alpha_1)-\kappa\cos(\kappa\alpha_1)-\kappa(\alpha_1-1)}
=\frac{\sin(\kappa(\alpha_1-1))}{\kappa(\alpha_1-1)}.
\end{equation}
Note that this equation is independent of $\beta_{1}$.
The right panel of Fig.~\ref{fig3} shows the divergence boundary \rf{divb} in the $(\alpha_1,\beta_1,\kappa)$-space.

The roots of the characteristic polynomial \rf{charp} are double imaginary if the discriminant of the biquadratic function vanishes:
\ba{discra}
&&(\sin(\kappa\alpha_1)-\kappa\alpha_1\cos(\kappa\alpha_1))^2(\tan\beta_1)^2\nn\\
&&+~2\alpha_1\kappa^2\tan\beta_1\cos\kappa\left[\cos(\kappa\alpha_1)+2(\alpha_1-1)\right]\nn\\
&&+~2\tan\beta_1\sin(\kappa\alpha_1)\left[2\sin(\kappa(\alpha_1-1))+\sin\kappa\right]\nn\\
&&-~\kappa\tan\beta_1\left[7\sin(\kappa(\alpha_1-1))(\alpha_1-1)
+\sin(\kappa(\alpha_1+1))(\alpha_1+1)\right]\nn\\
&&-~2\kappa\tan\beta_1\left[(2\alpha_1-3)\sin\kappa+\sin(\kappa(2\alpha_1-1))\right]\nn\\
&&+~(\sin\kappa-\kappa\cos\kappa)^2=0.
\ea
For this reason \cite{B1963,K2013,KS2004} equation \rf{discra} determines the boundary of the flutter domain that is shown in the left panel of Fig.~\ref{fig3}.

    \begin{figure}
    \begin{center}
    \includegraphics[angle=0, width=0.47\textwidth]{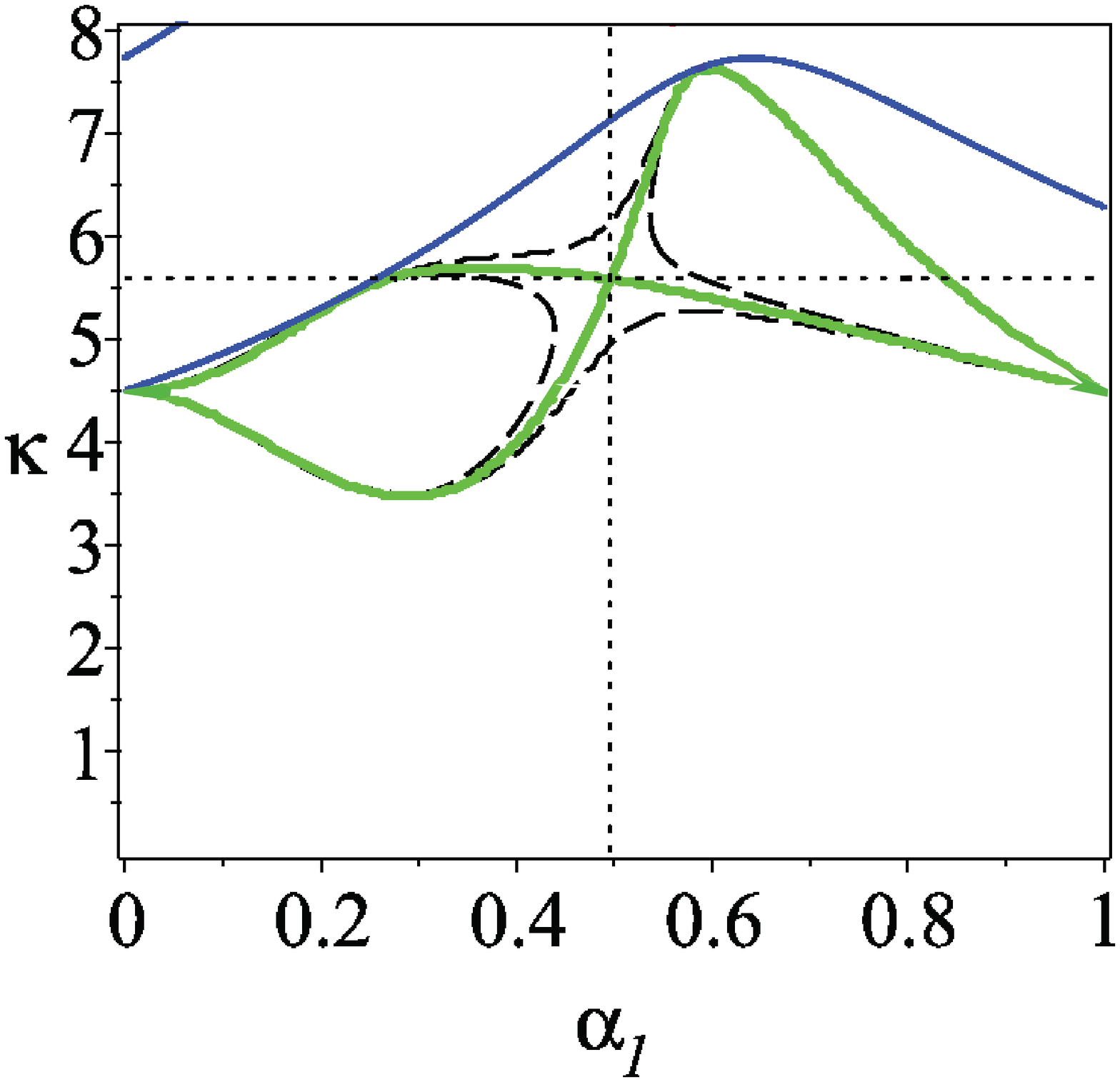}
    \includegraphics[angle=0, width=0.47\textwidth]{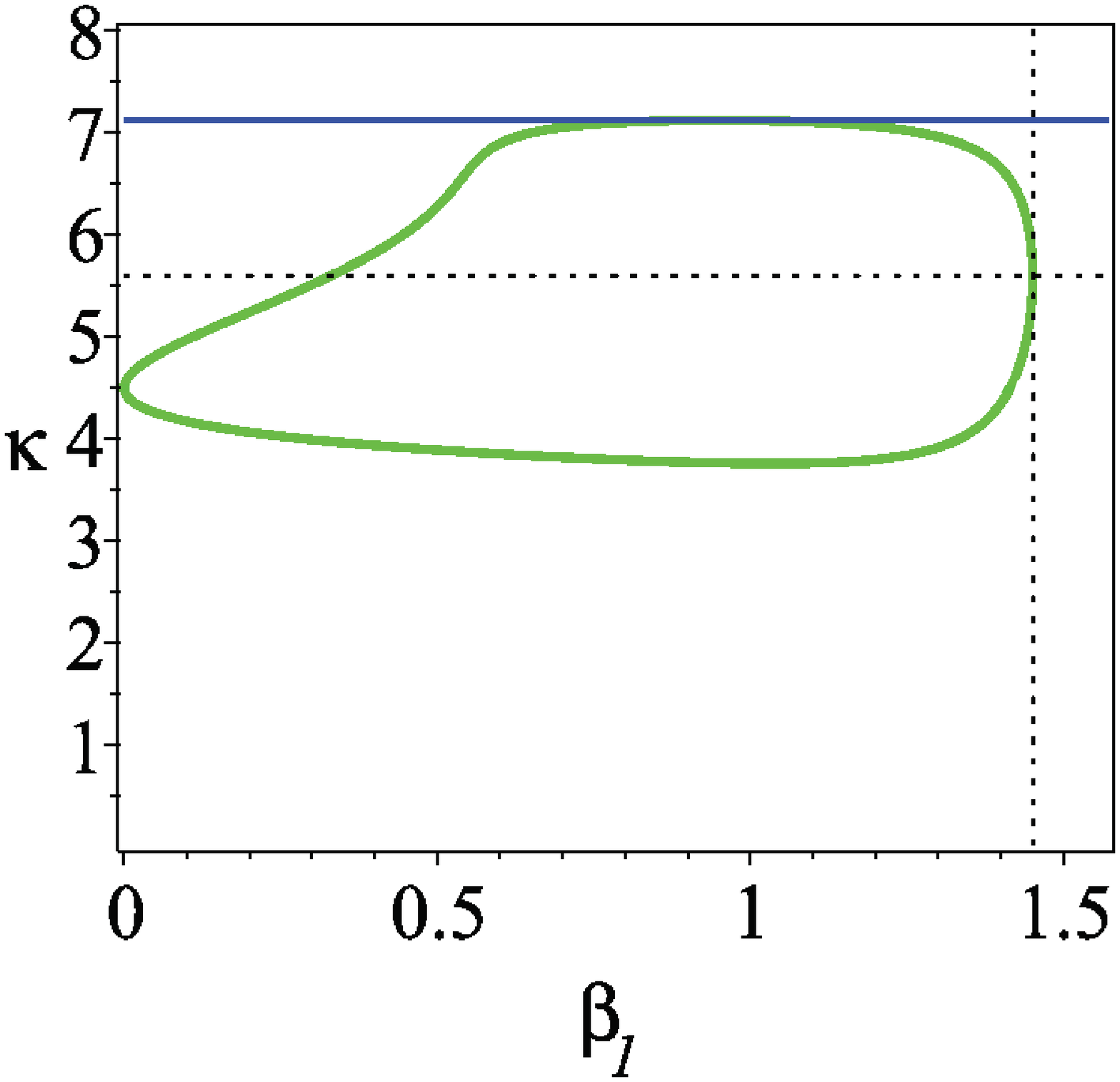}
    \end{center}
    \caption{Stability diagrams for (left) $\beta_{1}= {\hat\beta =} 1.450234089$ in the $(\alpha_1,\kappa)$-plane and (right) for $\alpha_1= {\hat\alpha =} 0.4947347666$ in the $(\beta_1,\kappa)$-plane. The solid blue curves designate the divergence boundary \rf{divb} and the solid green curves mark the flutter boundary \rf{discra}.
The flutter boundary in the left panel has a crossing at the saddle point located at  $\alpha_{1}=\hat\alpha$ and $\kappa = 5.591633160$.
The black dashed curves in the left panel correspond to the flutter boundaries at (upper and lower curves)
$\beta_1=\hat\beta-0.01$ and (left and right curves) $\beta_1=\hat\beta+0.01$.
In the right panel, the divergence boundary is a horizontal blue line with height $\kappa=\hat\kappa=7.113918994$.
 }
    \label{fig4}
    \end{figure}

For a given ($\alpha_1$, $\beta_1$), the critical value of the load parameter is given by
\[
         \kappacritalphaonebetaone = \min \{\kappa: (\kappa,\alpha_{1},\beta_{1}) \text{ satisfies either }\rf{divb} \text{ or }\rf{discra} \},
\]
as this is the length of the longest vertical line segment rising from the point $(\alpha_{1},\beta_{1},0)$ that does not enter either the flutter or the divergence domain.  Consequently, the quantity
\begineq{kappacritdef}
    \kappa^{*} = \sup\{\kappacritalphaonebetaone: \alpha_{1}\in[0,1],\beta_{1}\in[0,\pi/2)  \}
\end{equation}
is the supremum of all loads associated with a stable column.
Note that although the divergence boundary \rf{divb} is smooth, the boundary of the flutter domain \rf{discra} is nonsmooth.

Fig.~\ref{fig4} shows cross-sections of the flutter boundary and the divergence boundary
in the $(\alpha_1,\kappa)$- and $(\beta_1,\kappa)$-planes.
In the left panel, for which $\beta_1$ is fixed to $\hat\beta \approx 1.45$,
we see that the flutter boundary has a saddle point
in the $(\alpha_1,\kappa)$-plane at
$\alpha_1=\hat\alpha\approx0.495$, $\kappa\approx 5.59$.
On the other hand, when $\alpha_{1}$ is fixed to $\hat\alpha$, the flutter boundary has
a vertical tangent in the $(\beta_1,\kappa)$-plane at $\beta_1=\hat\beta$, as is visible in the right panel of Fig.~\ref{fig4}.
Consequently, when $\alpha_{1}=\hat\alpha$, the maximal
stable load $\kappacritalphaonebetaone$ varies smoothly for $\beta_{1}\in(0,\hat\beta)$, but when $\beta_{1}$ reaches $\hat\beta$
it jumps up discontinuously from the flutter boundary to the divergence boundary.
For the system under study such jumps were first described in the work \cite{KR2014} that corrected the classical result of Bolotin \cite{B1963}, whose plot in the $(\beta_1,\kappa)$-plane did not contain the divergence boundary at all, but provided a correct shape for the flutter boundary. Notice that
such overlapping of eigenvalue branches typically accompanies optimization of nonconservative systems and was reported in  numerous studies \cite{B1963,KS1998,Langthjem2012,SKK1976,C1975,HW1980,KK1980,BF2018,MB1981,LS2000cs,R1994,TF2007}. The general theory of this effect has been developed in \cite{K2013,KS2002pmm,KS2002vmgu}.

\begin{figure}
    \begin{center}
    \includegraphics[angle=0, width=0.47\textwidth]{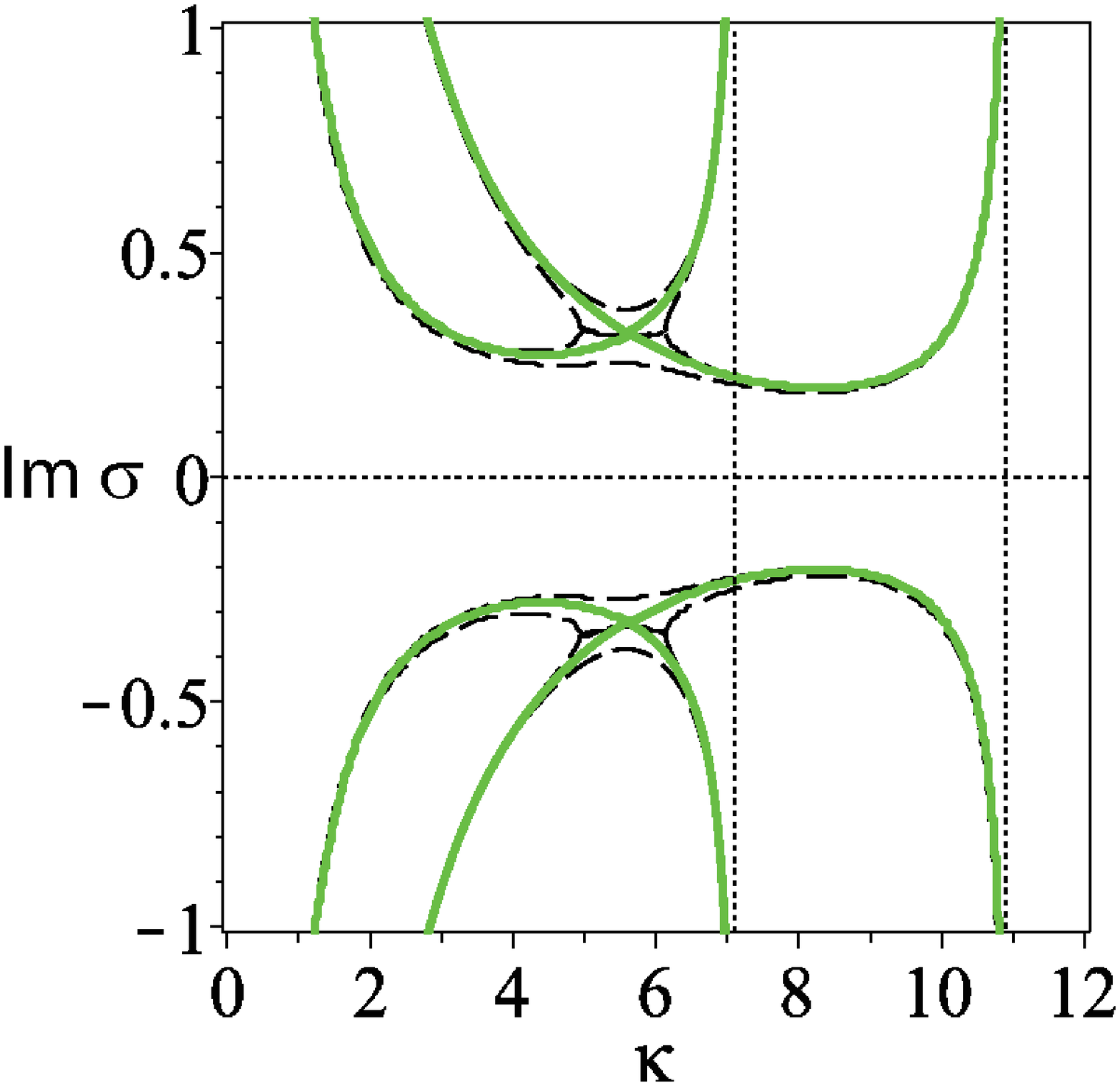}
    \includegraphics[angle=0, width=0.47\textwidth]{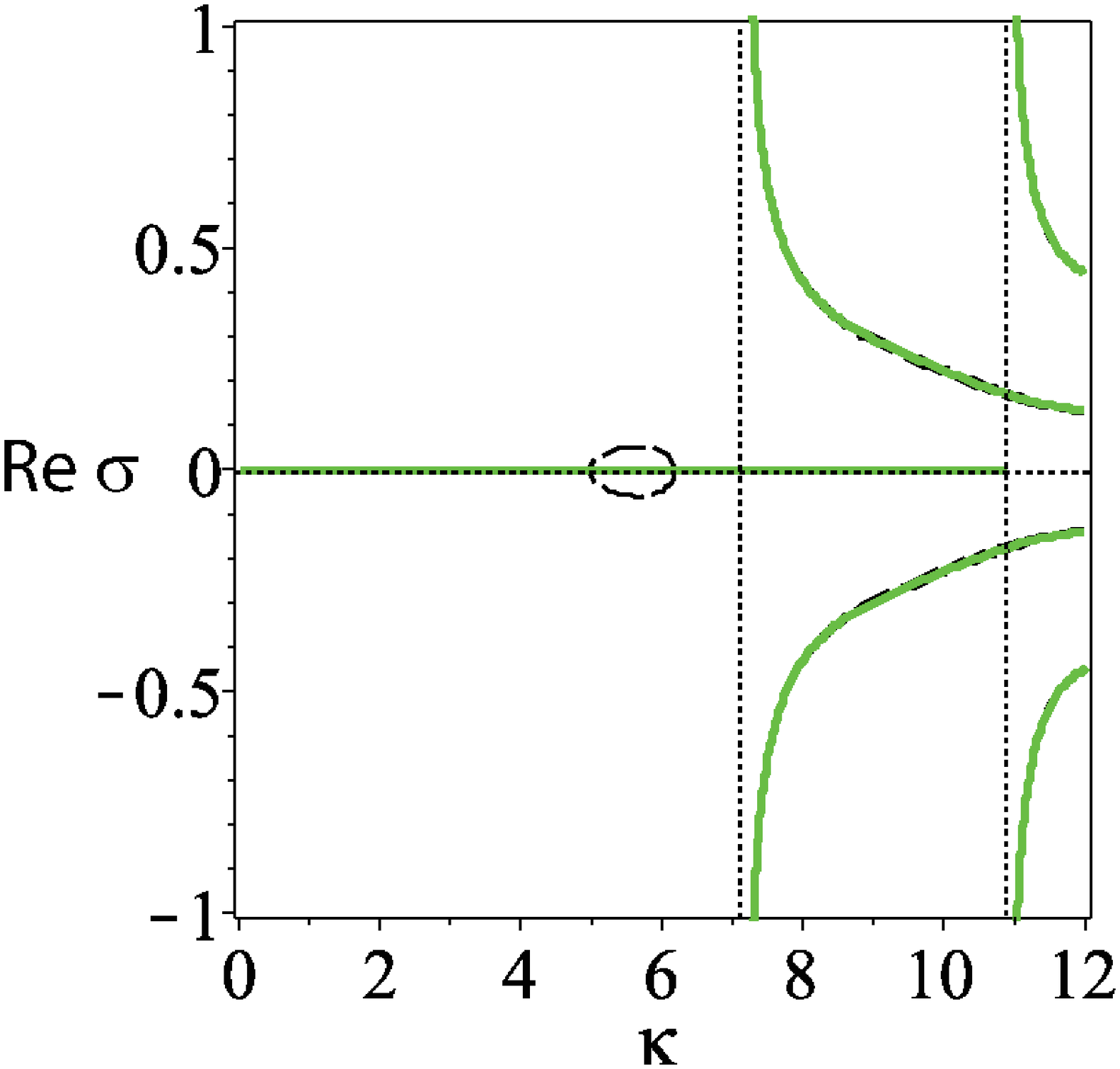}
    \end{center}
    \caption{(Left) imaginary and (right) real roots of the characteristic polynomial \rf{charp}
    for $\alpha_1=\hat\alpha=0.4947347666$ and (green, solid) $\beta_1=\hat\beta=1.450234089$ and (black, dashed)
      $\hat\beta\pm0.01$.
A bubble of complex eigenvalues appears for $\beta_1=1.450234089-0.01$ and corresponds to flutter instability. The black dotted vertical line at $\kappa=\hat\kappa=7.113918994$ is the onset of divergence instability. Increase in $\beta_1$ from
$\hat\beta-0.01$ to $\hat\beta+0.01$
results in the disappearance of the complex eigenvalues and hence
 is accompanied by the transition from the overlapping eigenvalue branches to an avoided crossing that yields a jump in the critical load parameter to the maximal value that is reached at $\kappa=\hat\kappa$ on the divergence boundary \cite{KR2014};
see also the right panel of Fig.~\ref{fig4}.
 }
    \label{fig5}
    \end{figure}

We can obtain a clearer picture of the jump discontinuity by plotting the real and imaginary parts of the eigenvalues $\sigma$
which describe the flutter boundary, as is done in Fig. \ref{fig5}. For $\alpha_{1}=\hat\alpha$, when
$\beta_{1}$ is decreased from the value $\hat\beta$, a bubble
of complex eigenvalues corresponding to flutter appears, but this vanishes for $\beta_{1}\geq\hat\beta$, resulting in the transition of
the critical load from the flutter boundary to the divergence boundary.

       \begin{figure}
    \begin{center}\
    \includegraphics[angle=0, width=0.47\textwidth]{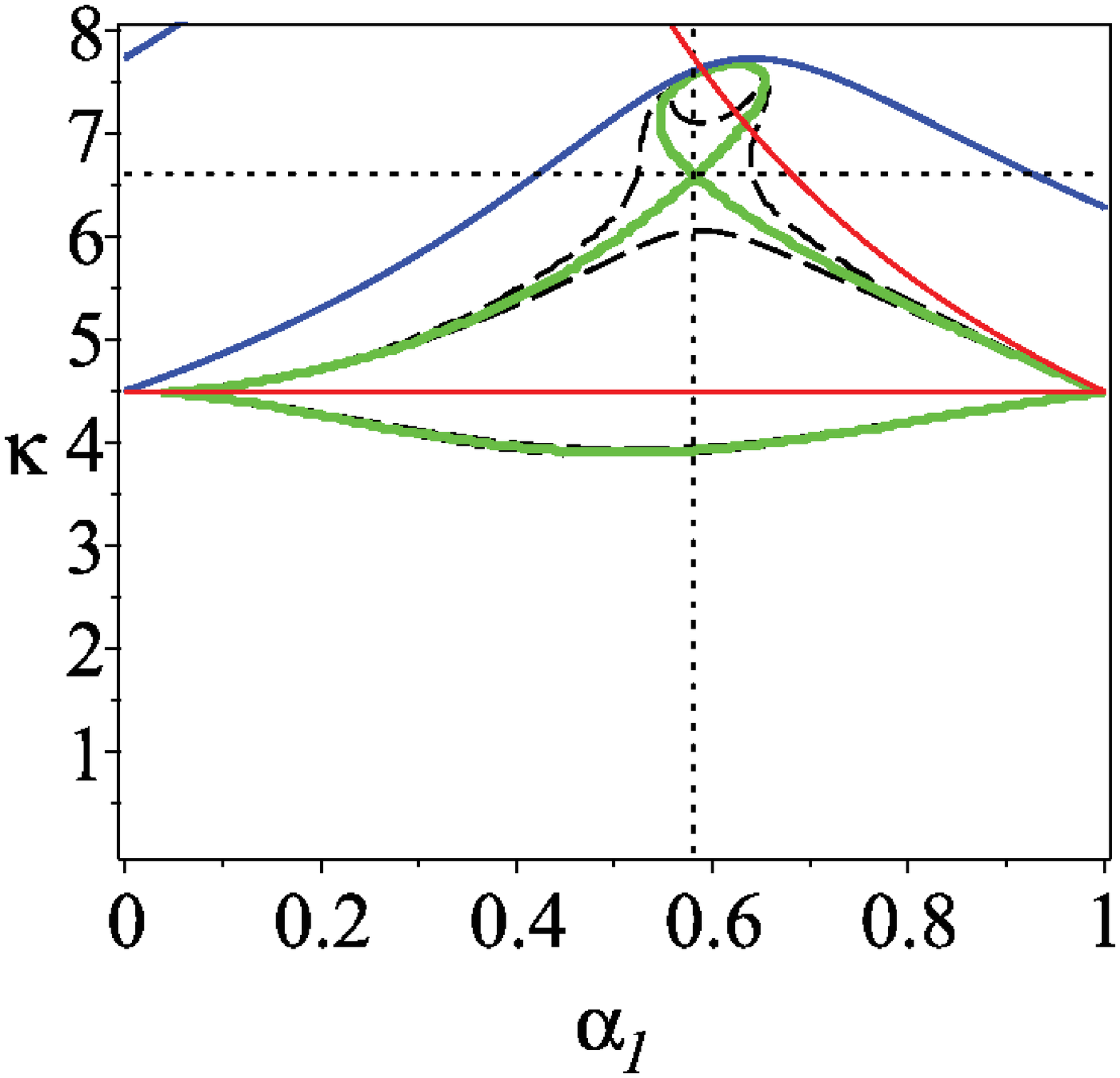}
    \includegraphics[angle=0, width=0.47\textwidth]{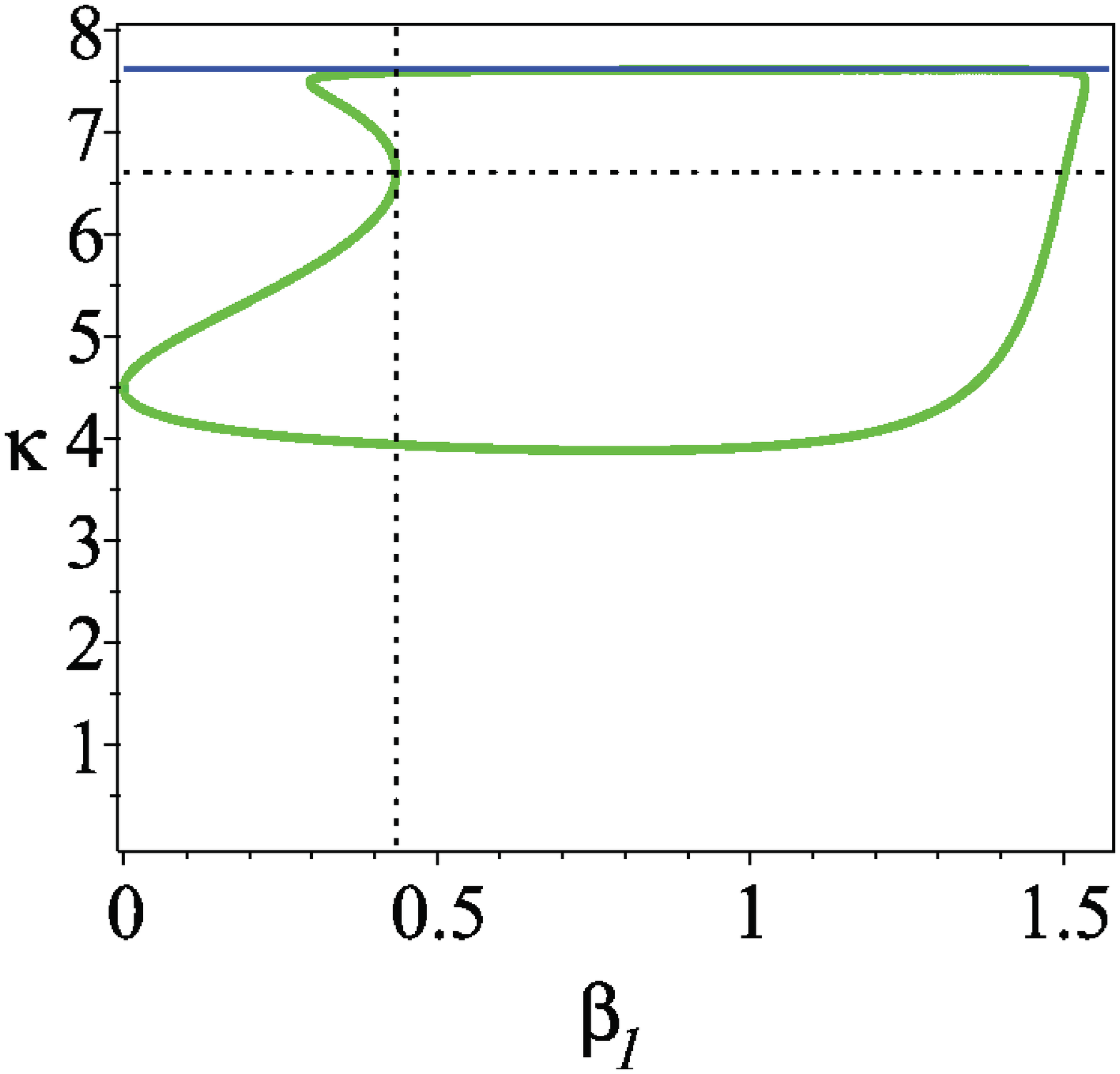}
    \end{center}
    \caption{Stability diagrams for (left) $\beta_1=\tilde\beta=0.4342999969$ in the $(\alpha_1,\kappa)$-plane and (right) for $\alpha_1=\tilde\alpha=0.5810701268$ in the $(\beta_1,\kappa)$-plane. The solid blue curves designate the divergence boundary \rf{divb}, and the solid green curves mark the flutter boundary \rf{discra}. The flutter boundary in the left panel has a crossing at the saddle point located at
$\alpha_{1}=\tilde\alpha$  and $\kappa = 6.600674669$. The black dashed curves in the left panel correspond to the flutter boundaries at (upper and lower curves)
$\beta_{1}=\tilde\beta-0.05$ and (left and right curves)
$\beta_{1}=\tilde\beta+0.05$.
    In the right panel, the divergence boundary is the horizontal blue line with height $\kappa=7.607584259$.
    The horizontal red line in the left panel shows the value $\kappa_{0}$ given in \eqref{kap0def}
     which is the smallest positive root of \rf{nomass}: the flutter boundary for the case $\beta_1=0$.
     The other red solid curve in the left panel is the solution to
   \rf{allmass}: the flutter boundary for the case $\beta_1=\pi/2$.
   }
    \label{fig6}
    \end{figure}

Looking at the discriminant \rf{discra} we notice that it degenerates into the equation
\begineq{nomass}
\kappa\cos(\kappa)-\sin(\kappa)=0
\end{equation}
for $\beta_1=0$ (i.e., when $\mu_1=0$) and reduces to the equation
\begineq{allmass}
\sin(\kappa\alpha_1)-\kappa\alpha_1\cos(\kappa\alpha_1)=0
\end{equation}
in the limit $\beta_{1}\to\pi/2$ (i.e., $\mu_{1}\to\infty$).
The sets defined by equations \rf{nomass} and \rf{allmass} are shown by the
solid red line and curve, respectively, in the left panel of Fig.~\ref{fig6}.
The flutter boundary is tangent to the planes $\beta_1=0$ and $\beta_1=\pi/2$ along
this line and curve. Note that the height of the red line is the smallest positive root of \rf{nomass}, which we denote by $\kappa_{0}$,
with
\begineq{kap0def}
          \kappa_{0}\approx 4.493409458.
\end{equation}
Since $\beta_{1}=0$ is the case where the mass $M_{1}=0$, $\kappa_{0}$ is the square root of the critical load
for the Dzhanelidze column (in view of \rf{dcp} and \rf{dck}).
%
%
%\mo{Let us define $\kappa_{0}$ to be the smallest positive root of \rf{nomass}: this is the critical load for the column
%in which the mass $M_{1}$ is zero (since $\beta_{1}=M_{1}/M_{2}=0). In the notation of Appendix \rf
%\begineq{kap0def}
%\kappa=\kappa_0 \approx 4.493409458,
%\end{equation}
%
%is the root of the equation \rf{nomass}, the corresponding set is just a straight line in   Fig.~\ref{fig6}.
%\mo{This might need rewriting: let's discuss: This is not surprising in view of the fact that,
%according to \rf{dcp} and \rf{dck}, $\kappa_0^2\approx 20.19$  which is the
%critical load for the Dzhanelidze column in which the relocatable mass $M_1$ is absent: $\mu_1=0$.}
The lines $\kappa=\kappa_0$ at $\alpha_1=0$ and $\alpha_1=1$ form singularities (edges) of the flutter domain.

           \begin{figure}
    \begin{center}
    \includegraphics[angle=0, width=0.47\textwidth]{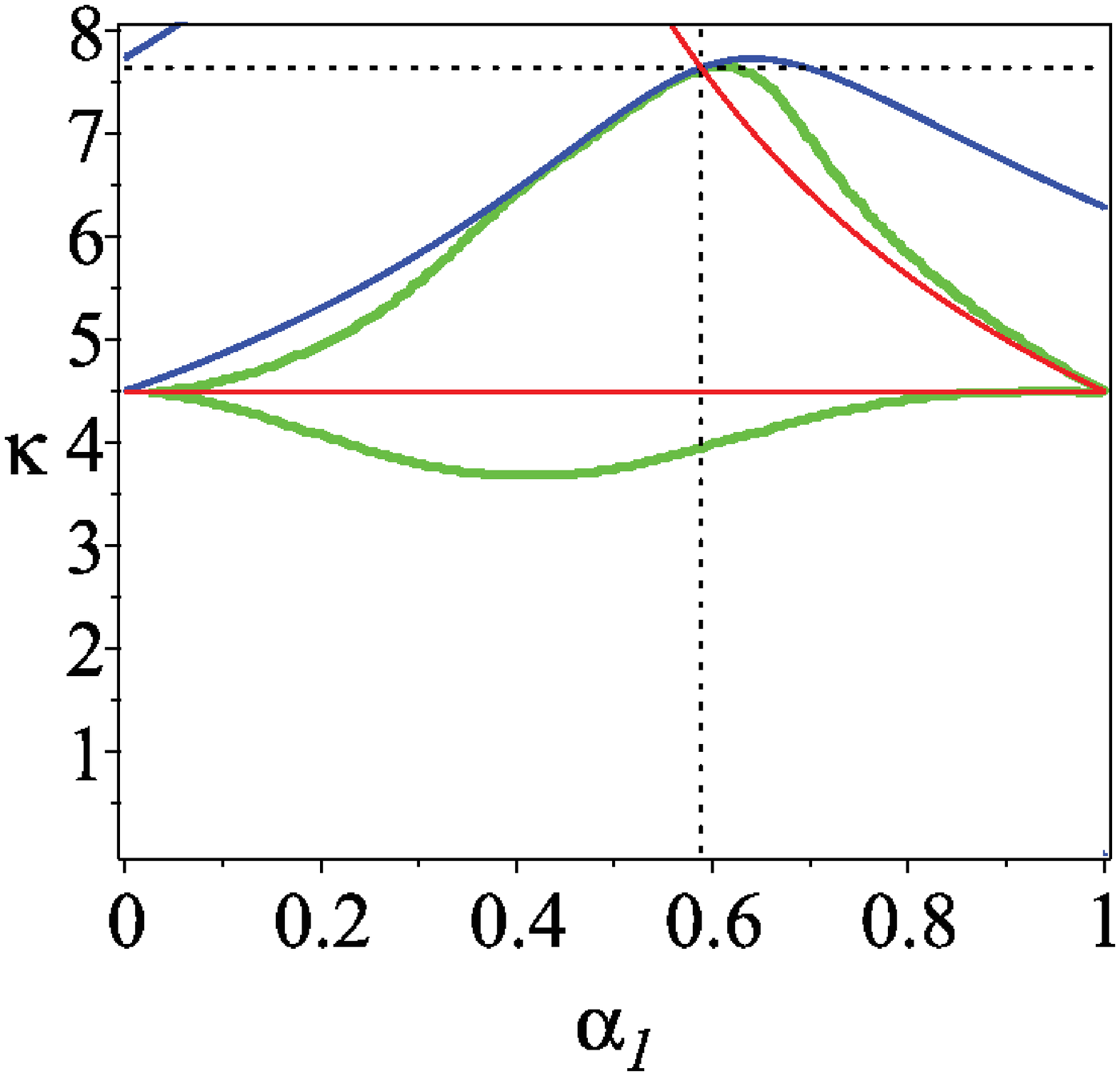}
    \includegraphics[angle=0, width=0.47\textwidth]{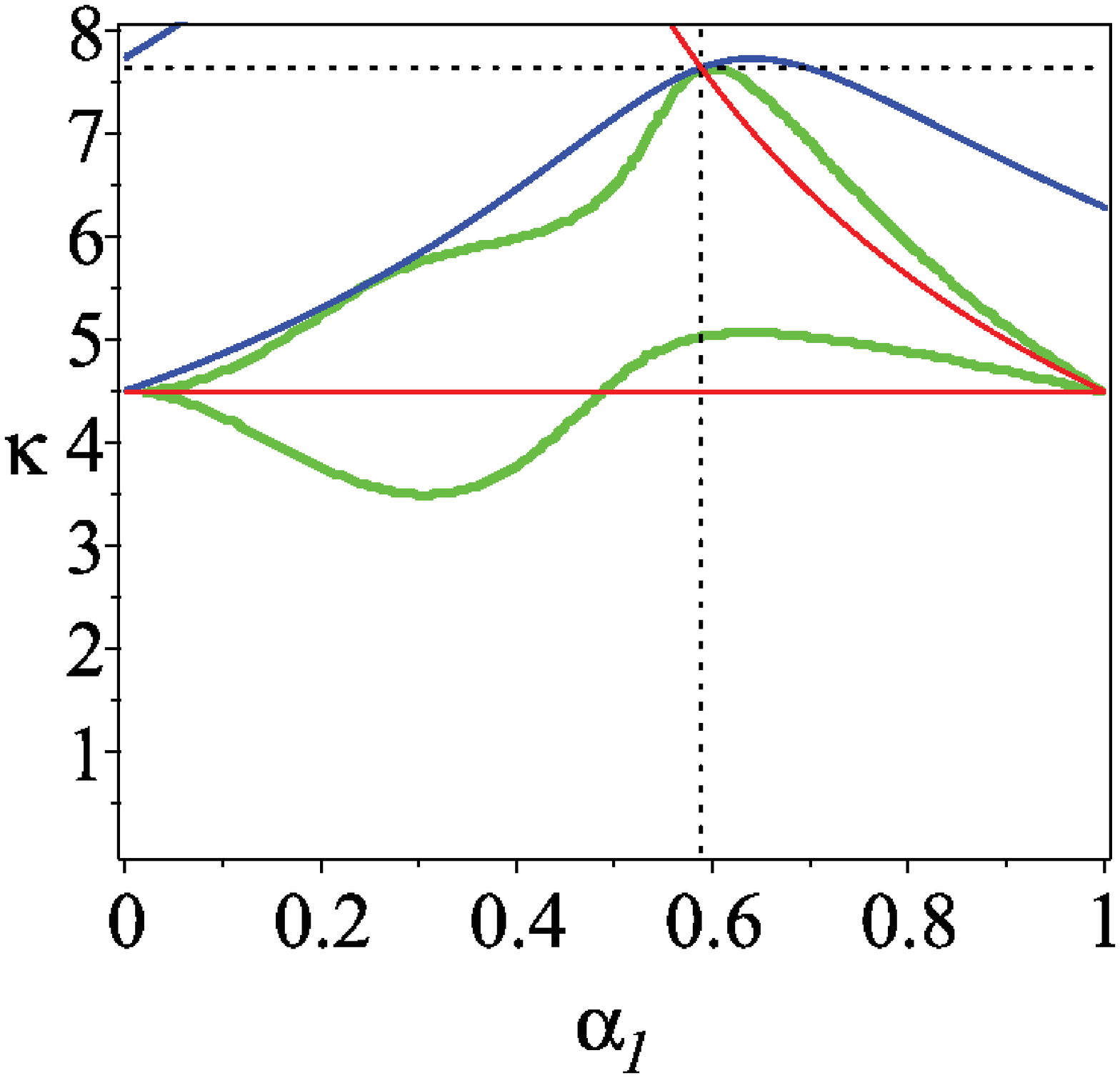}
    \includegraphics[angle=0, width=0.47\textwidth]{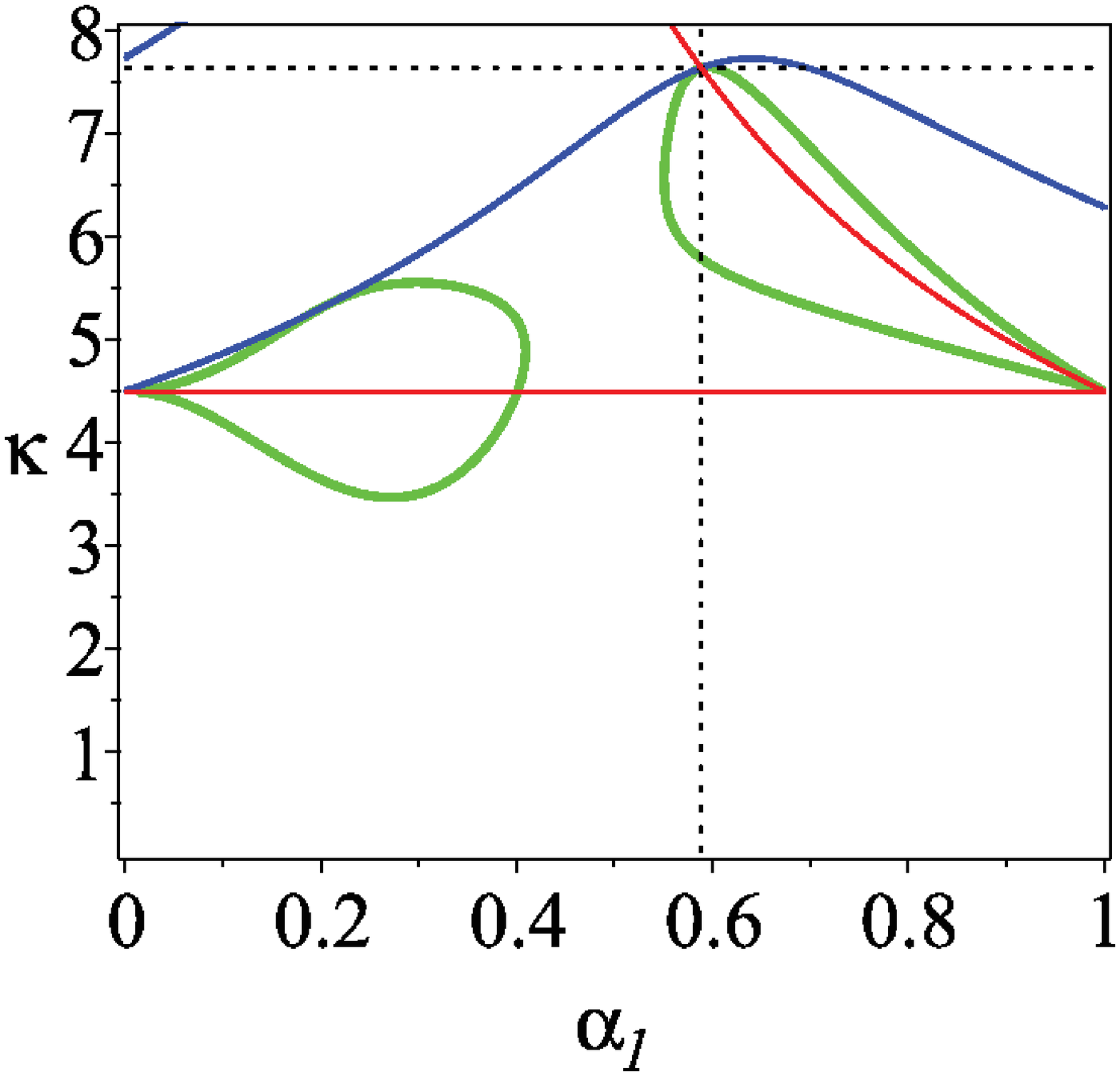}
    \includegraphics[angle=0, width=0.47\textwidth]{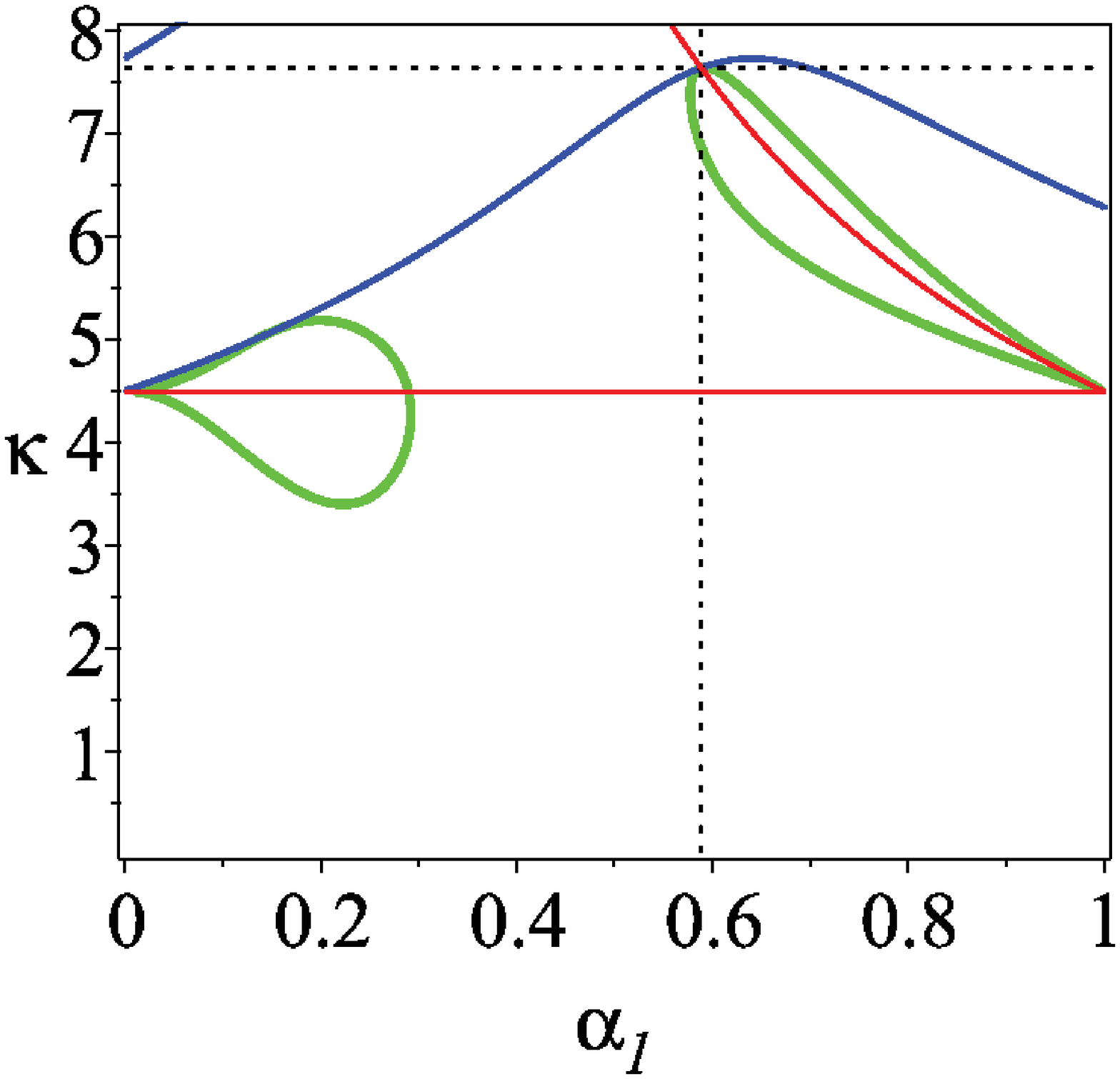}
    \includegraphics[angle=0, width=0.47\textwidth]{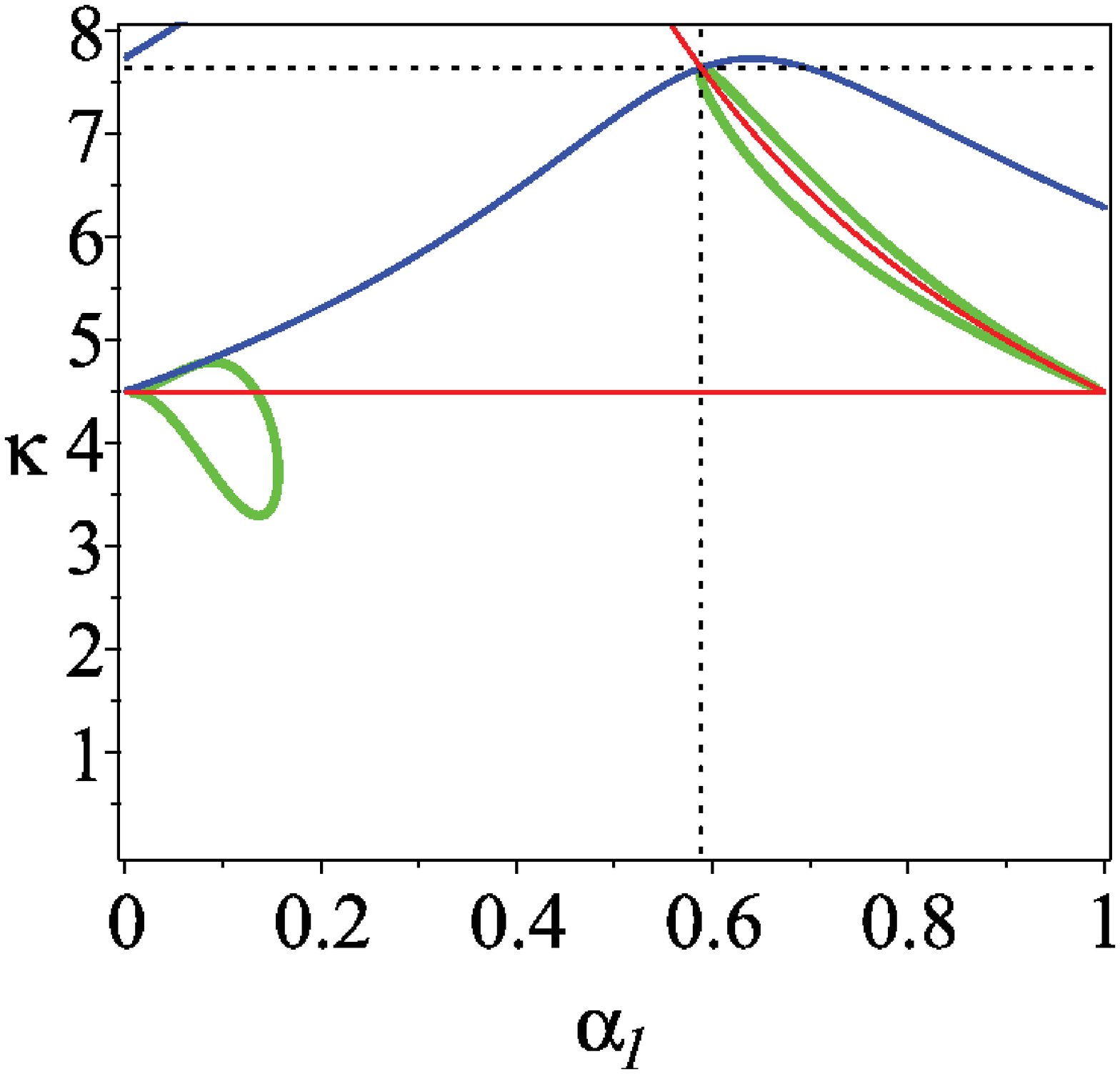}
    \includegraphics[angle=0, width=0.47\textwidth]{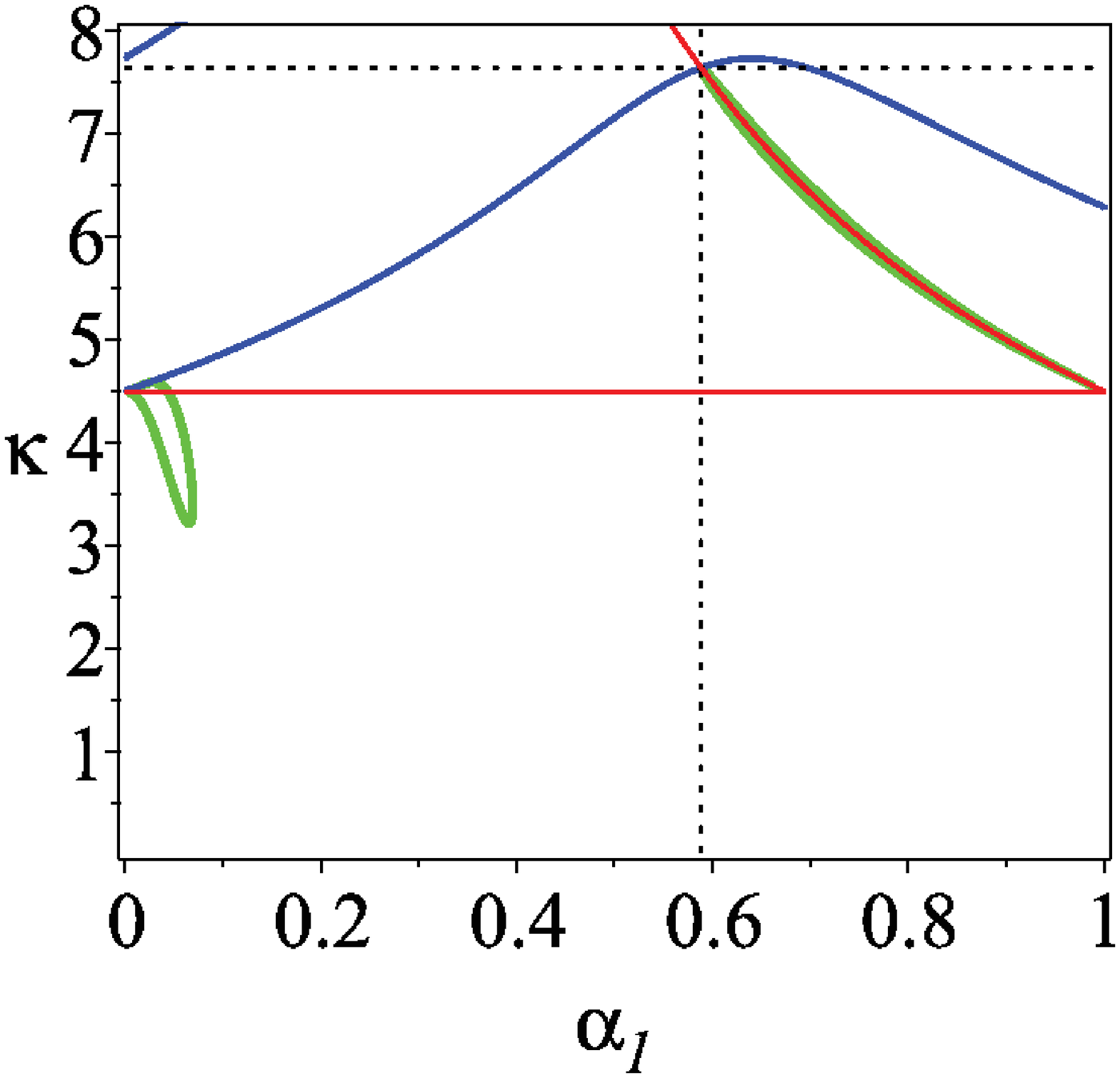}
    \end{center}
    \caption{Stability diagrams in the $(\alpha_1,\kappa)$-plane for (upper left) $\beta_1=\pi/2-0.5$, (upper right) $\beta_1=\pi/2-0.15$, (middle left) $\beta_1=\pi/2-0.1$, (middle right) $\beta_1=\pi/2-0.05$, (lower left) $\beta_1=\pi/2-0.01$, and (lower right) $\beta_1=\pi/2-0.001$. The black dashed lines intersect at the point with the coordinates of the optimal solution:  $\alphaopt \approx 0.588527598$ and $\kappaopt\approx 7.635002111$.}
    \label{fig7}
    \end{figure}

As soon as $\beta_1$ starts deviating from zero, a closed region of flutter instability appears around the horizontal red line $\kappa=\kappa_0$ in the $(\alpha_1,\kappa)$-plane. Furthermore, another region of flutter originates above it that touches the divergence boundary.
These two regions coalesce when $\beta_{1}$ reaches
$\tilde\beta\approx 0.434$;
see the left panel of Fig.~\ref{fig6}, which shows
another resulting saddle point on the flutter boundary defined by  \rf{discra}.
With further growth in $\beta_1$ the flutter region in the $(\alpha_1,\kappa)$-plane is simply connected, as shown in the two upper panels of Fig.~\ref{fig7} corresponding to $\beta_1=\pi/2-0.5$ and $\beta_1=\pi/2-0.15$, respectively, until this parameter passes the value $\beta_1 \approx 1.45$,
after which the flutter domain bifurcates into two parts; see the middle and the lower panels in Fig.~\ref{fig7}.

As $\beta_1$ approaches $\pi/2$, the upper portion of the flutter region concentrates around the red curve defined by \rf{allmass}, as shown
in the lower panels of Fig.~\ref{fig7}, and coincides with this curve exactly at $\beta_{1}=\pi/2$. At this very limit the critical load $\kappa$
reaches its supremal value $\kappaopt$, defined in \rf{kappacritdef}, which can be obtained by finding
the intersection point of the red curve defined by \rf{allmass} and blue curve defined by the divergence boundary \rf{divb}.
Solving the equations \rf{divb} and \rf{allmass} simultaneously, we find
\begineq{supremum-kappa-alpha}
\kappaopt\approx 7.635002112,\quad \alphaopt \approx 0.5885275986,
\end{equation}
and we write
\begineq{supremum-beta}
\betaopt=\frac{\pi}{2}
\end{equation}
to indicate that the supremum occurs in the limit $\beta_{1}\to\pi/2$.

    \begin{figure}
    \begin{center}
    \includegraphics[angle=0, width=0.31\textwidth]{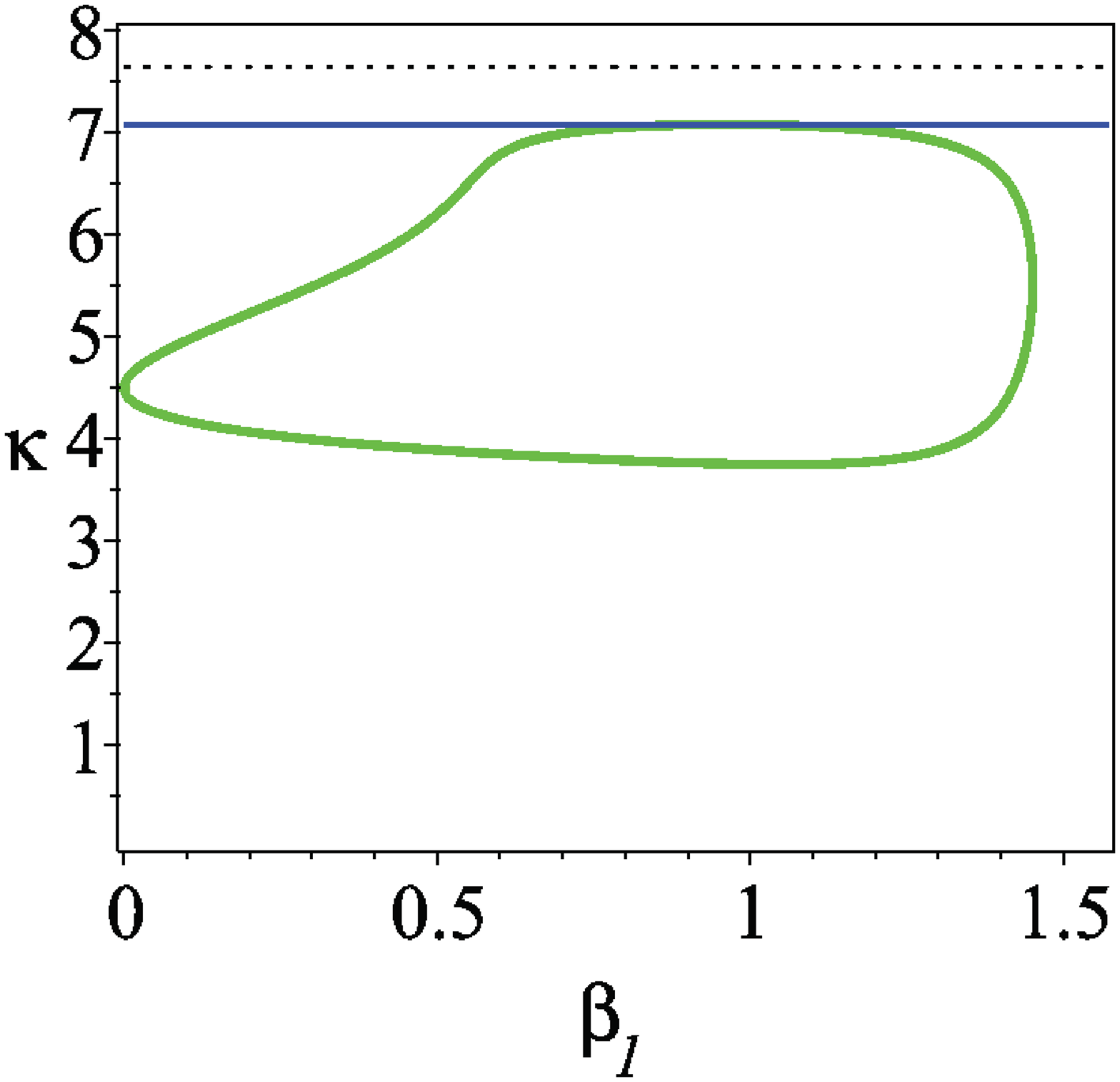}
    \includegraphics[angle=0, width=0.31\textwidth]{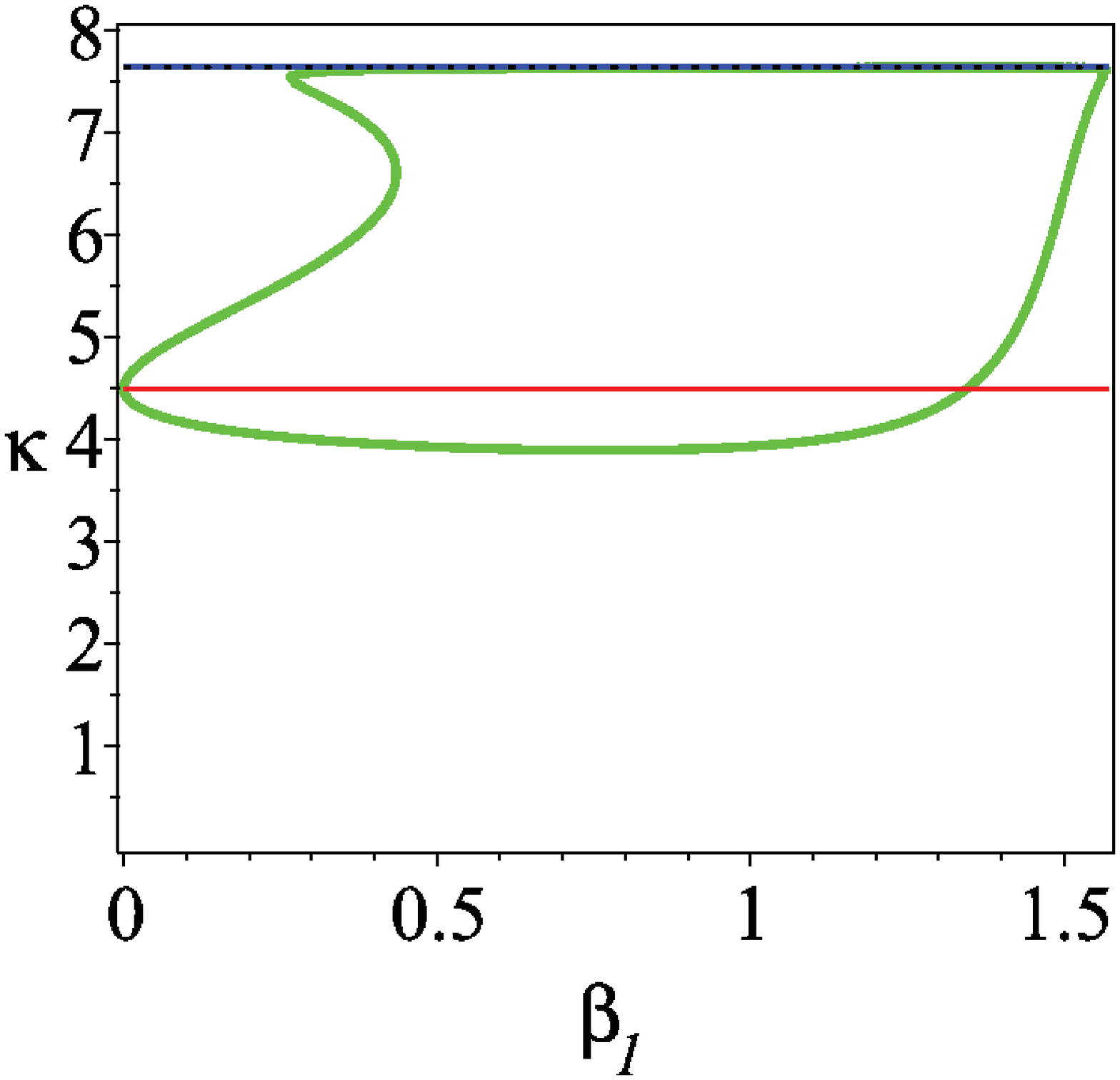}
    \includegraphics[angle=0, width=0.31\textwidth]{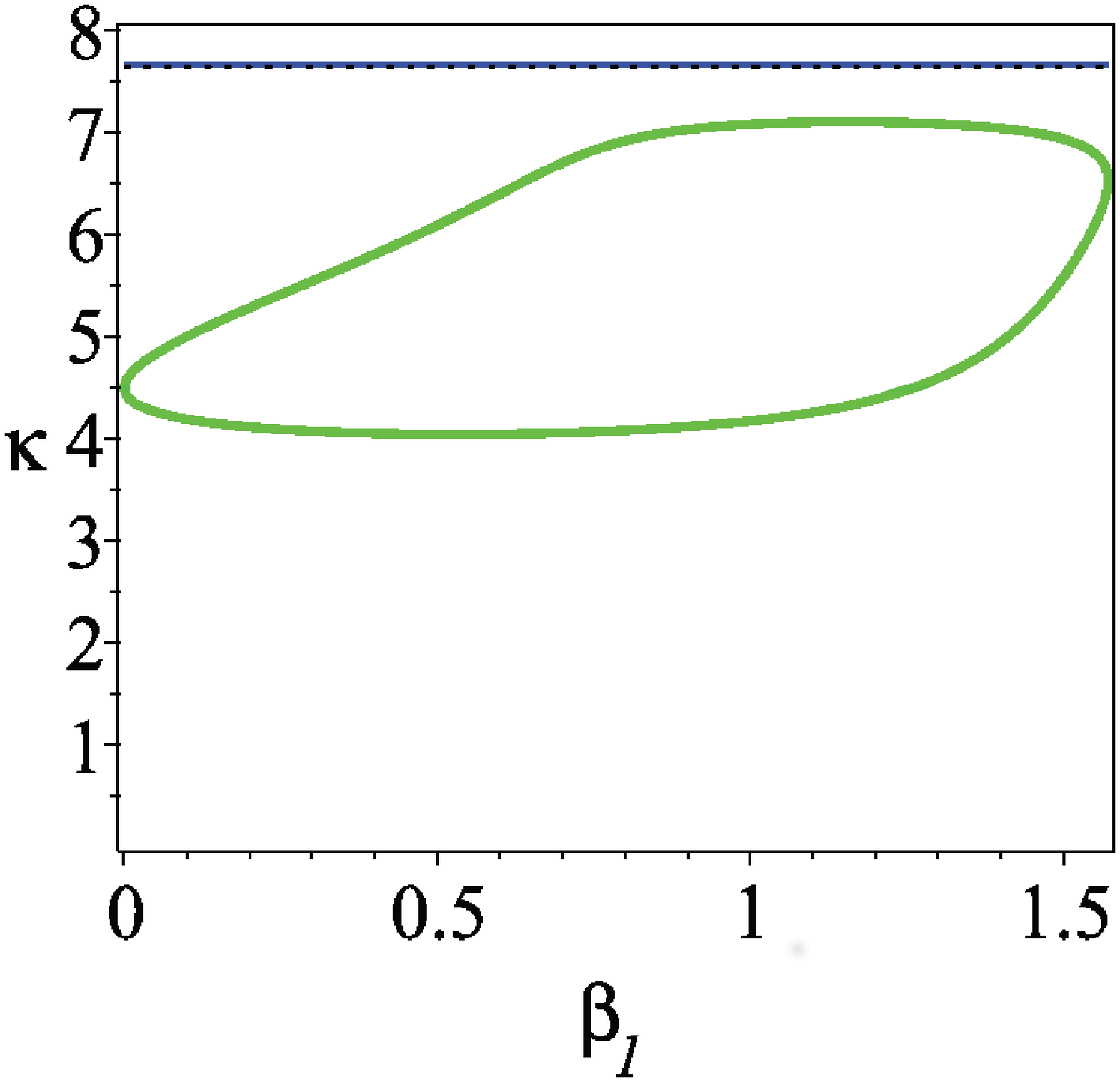}
    \end{center}
    \caption{
    Stability diagrams for (left) $\alpha_1=\alphaopt-0.1$ , (center) $\alpha_1=\alphaopt\approx 0.5885275986$  and (right) $\alpha_1=\alphaopt+0.1$. The green and blue curves respectively show the flutter and divergence boundaries. In the left and center panels,
    the critical load reaches the divergence boundary, but this is higher in the center panel, and there it is reached only if $\beta_{1}=\pi/2$.
    In the right panel, the flutter boundary prevents the critical load from reaching the divergence boundary.
    }

    \label{fig8} % formerly this was labelled fig9
    \end{figure}

Stability diagrams in Fig.~\ref{fig8} presented in the $(\beta_1,\kappa)$-plane show the decrease in the critical load $\kappa$ when $\alpha_1$ deviates from the value  $\alphaopt$,
indicating that the value $\kappaopt$ is a local supremum in the parameter space $\alpha_{1} \in [0,1]$,
$\beta_{1} \in [0, \pi/2)$.
Experiments reported in the next section strongly indicate that $\kappaopt$ is actually the global supremum.
However, note that $\M$ is not defined at $\betaopt=\pi/2$, since then the mass ratio $\mu_{1}=M_{1}/M_{2}$ is infinite,
so the supremum is not attained. Furthermore,
as $(\kappa,\alpha_{1},\beta_{1})\rightarrow (\kappaopt,\alphaopt,\pi/2)$, the matrix element $\M_{21}$ diverges to $\infty$ and
$\M_{12}$ converges to $0$ (see \rf{divb}), but $\M_{11}$
is the product of two quantities, one diverging to $\infty$ and the other converging to $0$ (see \rf{allmass}).
For this reason it is difficult to rigorously state limiting properties of the eigenvalues $\lambda_{k}$ of $\M$ as the supremum is approached,
though based on both our symbolic and numerical calculations,
it seems that, under the appropriate assumptions, the eigenvalues converge to a double zero eigenvalue with a Jordan
block, indicating that the parameters are on the boundary of both the flutter and divergence domains, and that the limiting
eigenvalues $\sigma_{k}$ of \rf{eip} coalesce into a quadruple eigenvalue at $\infty$.

 \begin{figure}
    \begin{center}
    \includegraphics[angle=0, width=0.45\textwidth]{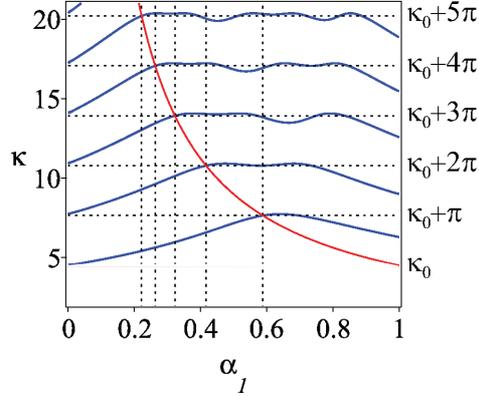}
    \end{center}
    \caption{Graphs of (red) equation \rf{allmass}  defining the flutter boundary in the limit $\beta_{1}\to\pi/2$
    as a function of $\alpha_{1}$ and  (blue) equation  \rf{divb} defining the divergence boundary as a function of $\alpha_{1}$.
     The intersection points are given by the expressions \rf{kapm} and \rf{kapm1}.
     }
    \label{fig9} % formerly this was labelled fig8
    \end{figure}

A key point in the derivation above is that the supremal value of 
$\kappacritalphabeta$ occurs when the
divergence boundary meets the flutter boundary in the limit $\beta_{1}\to\pi/2$.
We conjecture that this property holds for all $n$, not just for $n=2$.
If we substitute \rf{allmass}, which is the equation for
the flutter boundary in the limit $\beta_{1}\to\pi/2$, into the divergence boundary equation
 \rf{divb}, the latter can be simplified and reduced to
\begineq{simp}
\kappa(\alpha_1-1)\sin(\kappa(\alpha_1-1))(\cos(\kappa\alpha_1)-1)^2=0.
\end{equation}
Writing $\sin(\kappa(\alpha_1-1))=0$ yields
$\kappa\alpha_1-\kappa +  k \pi =0$,
$k\in \mathbb{Z}$. On the other hand, the relation \rf{allmass} can be written as $\kappa\alpha_1=\tan(\kappa\alpha_1)$, yielding $\kappa\alpha_1=\kappa_0$, where $\kappa_{0}$, given by \eqref{kap0def},
is the smallest positive root of the equation $\tan \kappa=\kappa$. Combining the results,
we obtain $\kappa=\kappa_{0} + \pi k,$ with $k\in \mathbb{Z}$.
For $k=0$, we obtain $\kappa=\kappa_{0}$,
 the optimal load when the mass $M_{1}$ is absent (and the square root of the critical load for
the  Dzhanelidze column), while for $k=1$, we obtain $\kappa=\kappa_0+\pi$, the supremum in \rf{supremum-kappa-alpha}
 just obtained for the optimal load for two concentrated masses $M_{1}$ and $M_{2}$. Let us therefore set $k=n-1$, giving
\begineq{kapm}
      \kappa=\kappa_{0} +  (n-1)\pi,
\end{equation}
and hence, using $\kappa\alpha_1=\kappa_0$,
\begineq{kapm1}
\alpha_1=\frac{\kappa_0}{\kappa_0+(n-1)\pi}.
\end{equation}
For $n=1$, the expression \rf{kapm1} yields $\alpha_1=1$, and for $n=2$, we have $\alpha_1=\kappa_0(\kappa_0+\pi)^{-1}$, which is the optimal value $\alphaopt$ given in \rf{supremum-kappa-alpha}.
This suggests a conjecture that \eqref{kapm} and \eqref{kapm1} are respectively the supremal value
$\kappa^{*}$ and the corresponding limiting value $\alpha_{1}^{*}$ for all $n$, with the corresponding limiting value $\beta_{1}^{*}$ equal to $\pi/2$.
Fig.~\ref{fig9} shows the values \rf{kapm} and \rf{kapm1} as defined
by the intersections of equations  \rf{divb} and \rf{allmass}, the divergence boundary equation and the flutter boundary equation
in the limit $\beta_{1}=\pi/2$, respectively. (It's perhaps worth noting that, for all $n$, we have
$\tan(\kappa_{0}+(n-1)\pi)=\tan(\kappa_{0})=\kappa_{0}$.)
%\begin{table}[ht]
%\centering
%\caption{The approximate values of the parameters $\kappa^{*}$ and $\alpha_1^{*}$ according to \rf{kapm} and \rf{kapm1}.}
%\label{tab1}
%\begin{small}
%\begin{tabular}{|c|c|c|}
%  \hline
%  % after \\: \hline or \cline{col1-col2} \cline{col3-col4} ...
%  $n$ & $\kappa^{*}=\kappa_0+(n-1)\pi$ & $\alpha_1^{*}=\kappa_0[\kappa_0+(n-1)\pi]^{-1}$ \\
%  \hline
%%  1 & $ 4.49340945790906$ &  1 \\
%%  2 & $ 7.63500211149886$ & $ 0.588527598589877$ \\
%%  3 & $ 10.7765947650886$ & $ 0.416960046829050$ \\
%%  4 & $ 13.9181874186784$ & $ 0.322844442508284$ \\
%%  5 & $ 17.0597800722682$ & $ 0.263391992093344$ \\
%%  6 & $ 20.2013727258580$ & $ 0.222430897092327$ \\
%  1 & $ 4.493$ &  1 \\
%  2 & $ 7.635$ & $ 0.5885$ \\
%  3 & $ 10.78$ & $ 0.4170$ \\
%  4 & $ 13.92$ & $ 0.3228$ \\
%  5 & $ 17.06$ & $ 0.2634$ \\
%  6 & $ 20.20$ & $ 0.2224$ \\
%  \hline
%\end{tabular}
%\end{small}
%\end{table}%

Remarkably, the numerical computations reported in the next section for $n$ concentrated masses, with $n=2,3,4,5$,
strongly indicate that the supremal load $\kappa^{*}$ and the corresponding limiting value $\alpha_{1}^{*}$ are
precisely the values given in \rf{kapm} and \rf{kapm1}
and illustrated in Fig.~\ref{fig9}, with the corresponding limiting value $\beta_{1}^{*}$ equal to $\pi/2$.
While we do not have conjectured formulas for the limiting values
$\alpha_{i}^{*}$ for $i>1$ and $n>2$, we conjecture that the corresponding limiting values $\beta_{i}^{*}$ are all $\pi/2$.
Indeed, the property $\beta_{1}^{*}=\pi/2$ implies that the mass ratio $\mu_{1}=M_{1}/M_{n}\to\infty$ as $\kappa\to\kappa^{*}$, which implies,
assuming that $M_{1}$ is bounded above, that $M_{n}\to 0$.  Consequently, if the other masses are nonzero
in the limit, all mass ratios $\mu_{i}=M_{i}/M_{n}$ must diverge to infinity as $\kappa\to\kappa^{*}$.

%%%%%%%%%%%%%%%% END OF THE THEORETICAL PART %%%%%%%%%%%%%%%%%%%%%%

\section{Numerical derivation of the optimal load for the massless column carrying multiple relocatable masses}\label{sec:experimental}

Recall that, as discussed in Section \ref{sec:weightless},
for a given number of masses $n$, our stability constraint is defined by
the eigenvalue problem $(\M\sigma^{2} + \K)u=0$ (see \rf{eip}).
Here $\K$ is the unit matrix while $\M$ is defined by \rf{deltaij} and \rf{mass}, which depend on the dimensionless
parameters $\alpha_{i}$ and $\mu_{i} = \tan\beta_{i}$, $i=1,\ldots,n-1$, defined in \rf{dlp},
as well as a given load $\kappa$.
Let us write $\M(\alpha,\beta,\kappa)$ for the matrix $\M$ defined by
 $\alpha=[\alpha_{1},\ldots,\alpha_{n-1}]^{T}$, $\beta=[\beta_{1},\ldots, \beta_{n-1}]^{T}$ and
$\kappa$. As noted in \rf{sigma-lambda}, the eigenvalues $\sigma_{k}$ of  $(\M(\alpha,\beta,\kappa)\sigma^{2} + \K)u=0$
are related to $\lambda_{k}$, the eigenvalues of the matrix
$\M(\alpha,\beta,\kappa)$, by $\sigma_{k} = \pm (-\lambda_{k}^{{-1}})^{{1/2}}$.

The stability constraint requires that, for given $(\alpha,\beta,\kappa)$,  all eigenvalues $\sigma_{k}$
should be imaginary, or equivalently, that all eigenvalue reciprocals $\lambda_{k}^{-1}$
are real and nonnegative.
Clearly, another equivalent condition is that all eigenvalues $\lambda_{k}$ are real and nonnegative,
interpreting $1/0$ as $+\infty$. Consequently, we define
a stability violation function $\tilde v: \R^{2n-1}\to\R_{+}$ by
\begineq{stabviolinit}
   \big (\alpha, \beta,\kappa \big ) \mapsto
              \max \left ( \Re \sqrt{-\lambda_{k} } \right ) ,
\end{equation}
where the maximum is taken over all eigenvalues of $\M(\alpha,\beta,\kappa)$, using the principal square root,
hence implying that  $\tilde v$ cannot take negative values.
Besides avoiding the nonlinearity in the reciprocal, the stability violation function $\tilde v$ has the virtue that it
is continuous, though not Lipschitz continuous, at points in parameter space where a positive eigenvalue $\lambda_{k}$
passes through the origin to the negative real axis, and hence
$\tilde v$  changes continuously from the value zero to a positive value that grows like the square root function at zero.  In this case, the parameters cross the divergence boundary,
since a
conjugate pair of imaginary eigenvalues $\sigma_{k}$ coalesce at $\infty$ and split along the real axis.
The function $\tilde v$ is also continuous, though not Lipschitz continuous, at points in
parameter space where two positive real eigenvalues $\lambda_{k},\lambda_{\ell}$ coalesce and split into a complex conjugate pair, and hence
again $\tilde v$ increases from zero to
a positive quantity that, generically, increases with the square root of the perturbation.
In this case, the parameters cross the flutter boundary, because
two simple imaginary eigenvalues $\sigma_{k},\sigma_{\ell}$ (and also their conjugates)
coalesce on the imaginary axis and split into a complex pair.

We argued in Section \ref{sec:analytical} that, in the case $n=2$, the optimal parameter configuration is simultaneously at
both the flutter boundary and the divergence boundary, likely with a double eigenvalue $\lambda$ at zero
(equivalently, a quadruple eigenvalue $\sigma$ at $\infty$) and, if this is the case, generically, the stability violation function
$\tilde v$ would grow at nearby parameter configurations with the fourth root of the perturbation.

To compensate for this non-Lipschitz behavior of $\tilde v$, we define a modified stability violation function $v: \R^{2n-1}\to\R$ by
\begineq{stabviol}
      v(\alpha,\beta,\kappa) = \left \{ \begin{array}{r} \tilde v(\alpha,\beta,\kappa)^{\rho}, \quad ~~~ \tilde v(\alpha,\beta,\kappa) \in [0,1]\\
                                                                                \rho \tilde v(\alpha,\beta,\kappa)- (\rho - 1), \quad ~ \tilde v(\alpha,\beta,\kappa) \in [1,\infty]
                                                       \end{array} \right .
\end{equation}
where $\rho$ is a positive integer. In the situations just discussed, the choice $\rho=2$ is sufficient to make
$v$ generically Lipschitz continuous at points where
the parameters cross either the divergence or the flutter boundary separately, and $\rho=4$ is sufficient to make
$v$ Lipschitz continuous even when the parameters cross the divergence and flutter boundaries simultaneously, at least
at the proposed optimal configuration \rf{supremum-kappa-alpha}, \rf{supremum-beta} for $n=2$.
In our computations, we experimented with choices of $\rho$ from 1 to 5 and we found that $\rho=4$ gave significantly
better results than $\rho<4$, but that setting $\rho=5$ made no further improvement.
Consequently, we chose to use $\rho=4$.
Note that the specific form of $v$ is
chosen so that it does not cause blow-up when $\tilde v(\alpha,\beta,\kappa)$ is large, and so that it is continuously differentiable
where $\tilde v(\alpha,\beta,\kappa)=1$.

However, what makes this problem particularly difficult is that as any $\beta_{i}\rightarrow\pi/2$,
the coefficient $\mu_{i}\rightarrow \infty $ in \rf{eip}. Consider the case $n=2$. We already mentioned
in Section \ref{sec:analytical} that as $\alpha_{1}\to\alphaopt$, $\beta_{1}\to \pi/2$ and $\kappa\to\kappaopt$, we have
$\M_{21}\to\infty$ and $\M_{12}\to 0$, while
$\M_{11}$ is a product of $\tan(\beta_{1})$ with a second factor that converges to zero. If this second factor
converges to zero more slowly than $(\tan(\beta_{1}))^{-1}$ does, so that $|\M_{11}|\to \infty$, a change
in its sign causes an eigenvalue $\lambda_{k}$ to discontinuously pass
through $\infty$ from the positive real to the negative real axis, implying infinitely large growth in the stability violation $v$
as the parameters cross the divergence boundary.
This presents a serious difficulty as we shall see.

In order to solve our optimization problem, we need to impose the stability constraint not only at a given point $(\alpha,\beta,\kappa)$,
but also at all points $(\alpha,\beta,\nu)$ with $\nu\in [0,\kappa]$. Although we could construct an approximation to $v(\alpha,\beta,\cdot)$
on the interval $[0,\kappa]$ using approximation software such as Chebfun \cite{DHT2014},
this is computationally expensive.
In our optimization computations, we found that a more effective approach is to impose the stability constraint on a coarse grid of $\tilde q$
logarithmically spaced points on $(0,\kappa]$, defining
\begineq{coarsegrid}
       c(\alpha,\beta,\kappa) = \max_{0\leq j\leq \tilde q} ( v(\alpha,\beta,\nu_{j}): \nu_{0}=\kappa, \nu_{j} = (1 - 2^{-j})\kappa, j=1,\ldots,\tilde q )
\end{equation}
and imposing the constraint $c(\alpha,\beta,\kappa) \leq 0$, or equivalently, $c(\alpha,\beta,\kappa) = 0$.
Then, after a potential solution is obtained by optimization, we check its stability on a much finer grid of $q\gg \tilde q$
uniformly spaced points on $(0,\kappa)$, rejecting it if this test is not passed.
We found that using a coarse grid with $\tilde q=10$ points and a fine grid with $q=10,000$ points worked well, typically with the majority
of the solutions obtained by optimization that are feasible for the coarse grid also passing the fine grid test.

We then pose our optimization problem as
\ba{optprob}
       \sup_{\alpha\in\R^{n-1},\beta\in\R^{n-1},\kappa\in\R} & \kappa \\
       {\rm subject~to~} & c(\alpha,\beta,\kappa) \leq 0, \nonumber \\
        &0 \leq \alpha_{1} \leq \ldots \leq \alpha_{n-1} \leq 1, \nonumber\\
        &0 \leq \beta_{i} \leq \pi/2,~ i=1,\ldots,n-1, \nonumber
\ea
This is not an easy problem to solve, since the stability constraint is nonconvex and nonsmooth, as well as discontinuous as $\beta_{i}\to\pi/2$.
We tackled it using \granso\ (GRadient-based Algorithm for Non-Smooth Optimization), a recently
developed open-source software package for nonsmooth constrained optimization \cite{CMO2017,Mit2020}.

As its name suggests, the algorithm implemented in \granso\ is based on employing user-supplied gradients.
This might seem contradictory since it is intended for nonsmooth
optimization problems, but although the constraints are not differentiable everywhere, they are differentiable
\emph{almost} everywhere.
Specifically, the stability violation function $v$ is differentiable at $(\alpha,\beta,\kappa)$ if the
following conditions hold:
\begin{enumerate} [(i)]
\item the maximum in \rf{coarsegrid} is attained only at one index $j \in (0,\ldots,\tilde q)$
\item the maximum in \rf{stabviolinit} is attained only at one eigenvalue $\lambda_{k}$ of $\M(\alpha,\beta,\nu_j)$
\item this eigenvalue $\lambda_{k}$ is simple and nonzero.
\end{enumerate}
Thus, evaluating the gradient of $v$ makes sense almost everywhere in parameter space. Of course,
the gradient does not vary continuously, but \granso\ is designed to exploit gradient difference information,
even near points where the gradient varies discontinuously, building a model of the constraint function on the parameter space using the Broyden-Fletcher-Goldfarb-Shanno (BFGS) quasi-Newton updating method.
For more details, see \cite{CMO2017}, and for application of BFGS in other stability optimization problems, see \cite{Ove2014} and
the papers cited there.

To derive the gradient of $v$, we need to differentiate an eigenvalue $\lambda_{k}$ with respect to changes in the matrix $\M$.
Let us write $\M(t)=\M + t({\bf \Delta}\M)$ and let $\lambda(t)$ denote the eigenvalues of $\M(t)$.
It is well known \cite{GLO2020} that, if $\lambda_{k} = \lambda(0)$ is a simple eigenvalue of $\M=\M(0)$ satisfying the right and left eigenvector
equations $\M u=\lambda u$ and $w^{*} \M^{*} = \lambda w^{*}$, where the asterisk denotes complex conjugate transpose, then
\[
                \left .  \frac{d}{dt} \lambda(t)\right |_{t=0} = \frac{w^{*}( {\bf \Delta}\M )u} {w^{*} u}.
\]
With this in mind, deriving the gradient of $v$ with respect to the $2n-1$ parameters given by $(\alpha,\beta,\kappa)$ is
straightforward, employing the chain rule to incorporate the variation in the power function  in \rf{stabviol}, the square root in \eqref{stabviolinit},
and the formulas \rf{mass}, \rf{deltaij} and \rf{tanbeta}.

We now describe our experiments using \granso\ (version 1.6.4), running in \matlab\ (release R2020a) on a MacBook Air laptop,
to solve \rf{optprob}. We used the default choice of parameters with the following
exceptions: we set {\tt maxit}, the limit on the iteration count, to 500,
and we set the tolerances \verb@opt_tol@ and \verb@feas_tol@ to zero, to obtain the highest possible accuracy.
We added bound constraints on the load variable
formulated as $0 \leq \kappa \leq \kmax$ with $\kmax=1.1\times (\kappa_0+(n-1)\pi)$, that is, with a lower bound of zero and an upper
bound set to 10\% higher than the proposed optimal value of $\kappa$ given in
\rf{kapm}.
Since \granso\ may generate iterates violating these bounds or the other bound constraints in \eqref{optprob}, we defined $v$ to be zero if
$\kappa\leq 0$ and replaced $\beta_{i}$ in \rf{tanbeta} by \verb@pi/2@, the 16 digit rounded value of $\pi/2$,
if $\beta_{i}$ exceeds \verb@pi/2@, to avoid the discontinuity in the tangent function at $\pi/2$
(note that \verb@tan(pi/2)@ $\approx 1.6\times 10^{16}$ has the desired positive sign). Because of the difficulty of the problem,
we ran the code from many randomly generated starting points, with the initial values for
$\kappa$, $\alpha_{i}$ and $\beta_{i}$ generated from the uniform distribution on $[0,\kmax]$, $[0,1]$ and $[0,\pi/2]$ respectively,
with the $\alpha_{i}$ then sorted into increasing order.

\begin{figure}[t]
\begin{center}
 % left bottom right top : need a bit less cut off the right as otherwise title is chopped
 \includegraphics[scale=0.5,trim={1.5cm 1.5cm 1.3cm 1.5cm},clip]{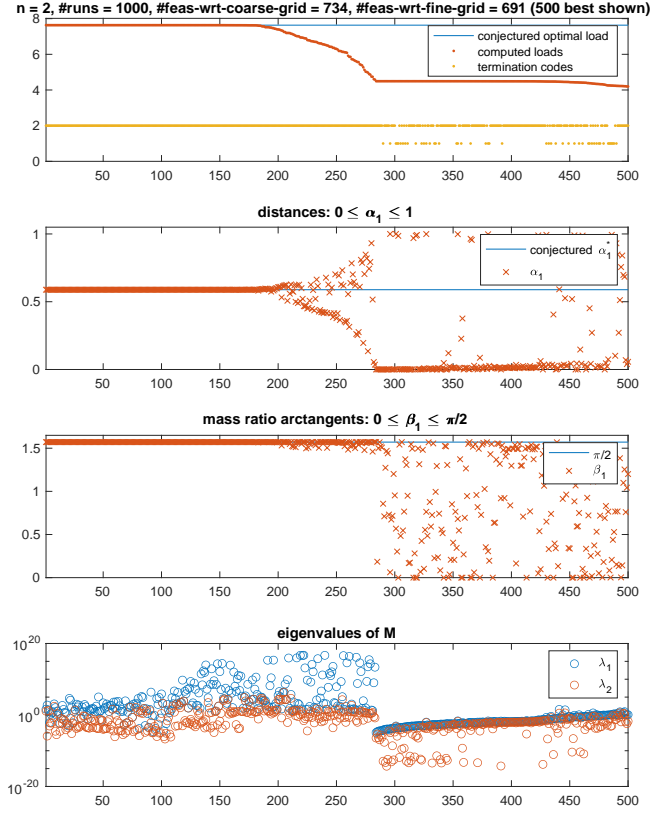}
 \end{center}
 \caption{Summary of results for solving \eqref{optprob} with $n=2$,
running \granso\ from 1000 randomly generated starting points. Of the 1000 candidate solutions
obtained, 734 satisfied the bound and coarse grid stability constraints imposed by \granso,
and of these, 691 also passed the fine grid stability test. The top panel in the figure shows the computed optimal
loads $\kappa$ for the best 500 of these feasible solutions, sorted into decreasing order; the top 100 final values all agree with
 $\kappa_{0}+\pi$ to 4 digits, while the top two final values agree with $\kappa_{0} + \pi$ to ten digits.
The second and third panels show the
associated final values of $\alpha_{1}$ and $\beta_{1}$ computed by these same 500 runs. The fourth panel shows the eigenvalues
of the final associated matrix $\M(\alpha_{1},\beta_{1},\kappa)$. The computed solutions clearly separate into four
flavours; see the text for details.
}
 \label{fig:n2Beta1LimPiOver2}
 \end{figure}

\subsection{Results for $n=2$}

Our analytical discussion of the case $n=2$ was given in Section \ref{sec:analytical}; the results here strongly support our
claim that the optimal configuration is given by \rf{supremum-kappa-alpha}, \rf{supremum-beta}.
Fig.\ \ref{fig:n2Beta1LimPiOver2} shows the results obtained by
running \granso\ from 1000 randomly generated starting points. Of the 1000 candidate solutions
generated by \granso, 734 satisfied the bound and coarse grid stability constraints imposed by \granso,
and of these, 691 also passed the fine grid stability test described above. The top panel in the figure shows the computed optimal
loads $\kappa$ for the best 500 of these feasible solutions, sorted into decreasing order, while the second and third panels show the
associated final values of $\alpha_{1}$ and $\beta_{1}$ computed by these same 500 runs. The fourth panel shows the eigenvalues
of the final associated matrix $\M(\alpha_{1},\beta_{1},\kappa)$.
The highest two final values of $\kappa$ agree with each other, and with the
value $\kappa_0 + \pi$ given in \rf{supremum-kappa-alpha} and \rf{kapm},
to 10 digits. The final values for $\alpha_{1}$ and $\beta_{1}$ for these same two best results
agree with the value $\kappa_0[\kappa_0+\pi]^{-1}$ (given in \rf{supremum-kappa-alpha}
and \rf{kapm1}) and $\pi/2$, to 10 and 12 digits,
respectively. It's also worth noting  that the top 100 final values for the computed optimal load agree with $\kappa_{0}+\pi$ to 4 digits.

Looking at all four panels of Fig.\ \ref{fig:n2Beta1LimPiOver2}, we see that the top 500 results come in several clearly distinct flavours.
The first flavour is exhibited by the best 180 or so runs which all give good approximations to $\kappa_{0}+\pi$.
However, starting with the 284th result, we find a very different second flavour: many runs find that the computed optimal load is about $\kappa=4.493$,
which agrees with $\kappa_{0}$, the square root of the critical load for the Dzhanelidze column, to four digits.
Clearly, this is a locally maximal value for \rf{optprob}; otherwise, it would not be found so frequently.
If we look at the associated computed $\alpha_{1}$ and $\beta_{1}$ values, usually $\alpha_{1}$  is close to zero, but if not,
then $\beta_{1}$ is close to zero. It is easily checked that, regardless of the value of $\beta_{1}$, if $\alpha_{1}=0$ then
$\M(\alpha_{1},\beta,\kappa_{0})$ is the zero matrix, with a double semisimple zero eigenvalue, so this locally optimal
parameter configuration, like the apparent globally optimal configuration \rf{supremum-kappa-alpha}, \rf{supremum-beta},
is on both the flutter and divergence boundaries.
Physically, this corresponds to the mass $M_{1}$ being fixed at the clamped end of the column.
On the other hand, regardless of the value of $\alpha_{1}$, if $\beta_{1}=0$, then $\M(\alpha_{1},\beta_{1},\kappa_{0})$
has all zero entries except for $\M_{12}$, and hence has a double zero eigenvalue with a Jordan block.
Again, this parameter configuration is on both the flutter and divergence boundaries, and physically, it corresponds to the mass $M_{1}$ being zero.
Note that the computed eigenvalues for this second flavour of solutions are relatively small.

\begin{figure}[t]
\begin{center}
\begin{tabular}{cc}
 \includegraphics[scale=0.39,trim={1.5cm 1.5cm 1.3cm 1.5cm},clip]{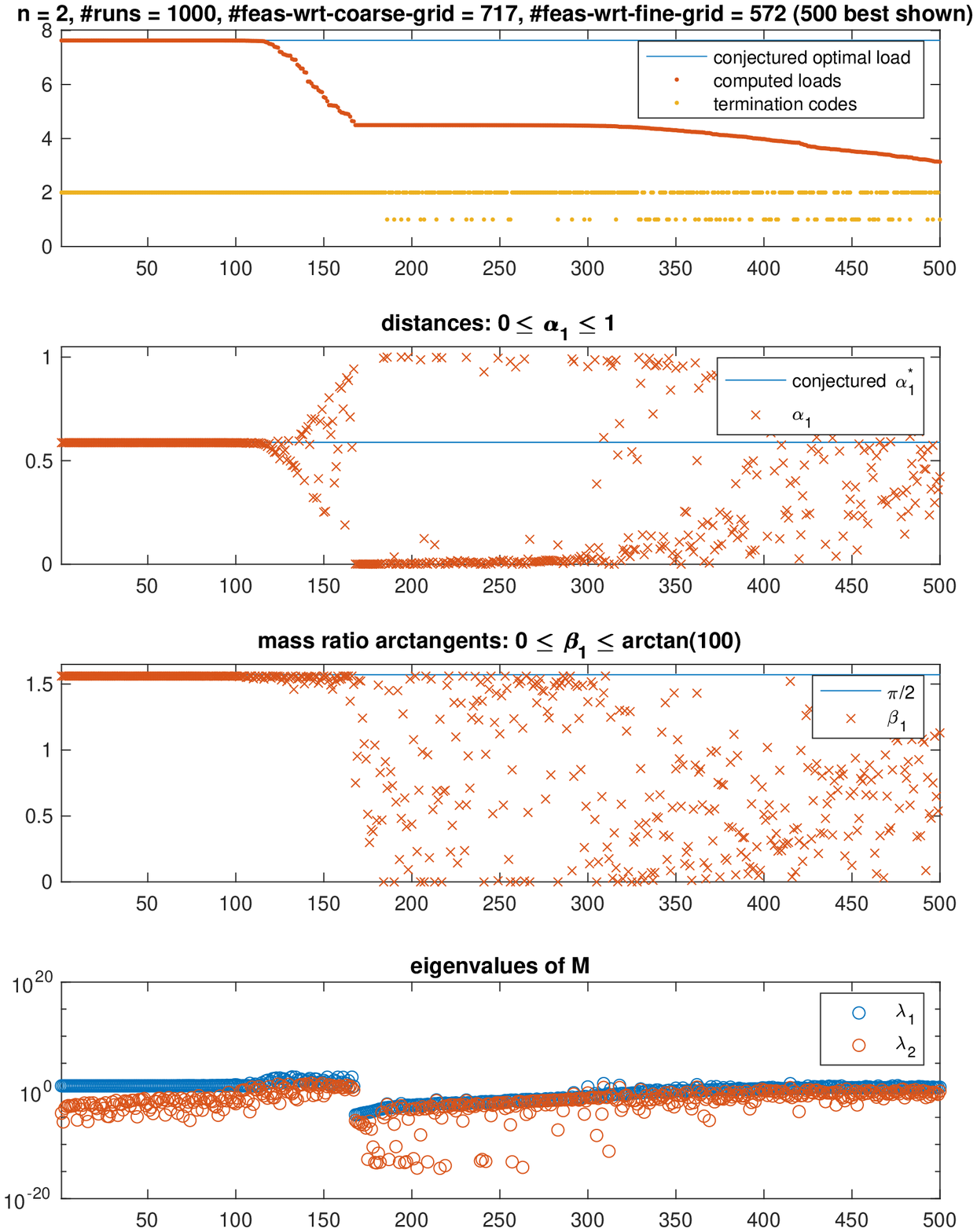} &
 \includegraphics[scale=0.39,trim={1.5cm 1.5cm 1.3cm 1.5cm},clip]{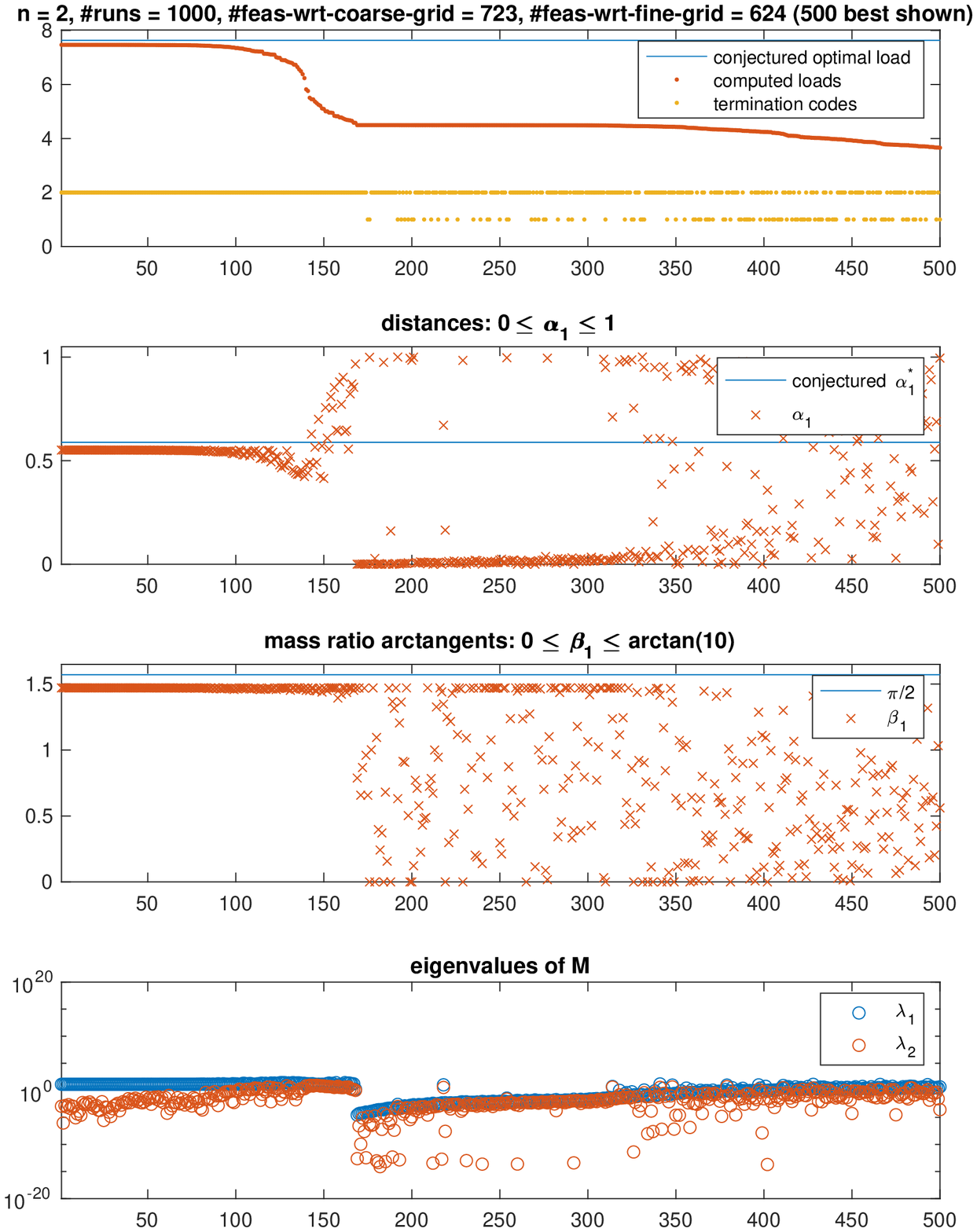}
 \end{tabular}
 \end{center}
 \caption{Solving \eqref{optprob} for $n=2$ with the additional constraint (left) $\beta_{1}\leq \arctan(100)$ and (right)
$\beta_{1}\leq \arctan(10)$. The best computed optimal loads are
respectively just  0.1\%  and 2\% lower than the apparent unconstrained supremum $\kappa_{0}+\pi$. Note the
dramatic reduction in the size of the eigenvalues of the final computed $\M$ compared to Fig.\ \ref{fig:n2Beta1LimPiOver2}.
See the caption of Fig.\ \ref{fig:n2Beta1LimPiOver2} and the accompanying text for more details.}
 \label{fig:n2Beta1LimArctan100-10}
\end{figure}

The third flavour of results is exhibited by the results numbered approximately 180 to 280. In these cases, {\granso} terminated
prematurely, without approximating a globally or locally maximal value, and we can see also that, on average, the larger $\kappa$ is, the closer
$\alpha_{1}$ is to $\kappa_0[\kappa_0+\pi]^{-1}$.  Investigation of these cases shows that termination occurs
because of the discontinuity in the stability constraint that we described above. This is also supported by the enormous associated eigenvalues
of $\M$ shown in the fourth panel. Note also that as $\kappa$ increases
towards its optimal value, these eigenvalues decrease, but they neither converge to specific values, nor do they become very small.
In fact, the matrix $\M$ associated with the best computed optimal $\kappa$ is
\[ \footnotesize
\begin{bmatrix}  5.4382\times 10^{1}  & -4.0893\times 10^{-10}\\
   1.3020\times 10^{12}  & -6.8246 \times 10^{-1}
\end{bmatrix}
\]
\normalsize
Although its eigenvalues are not close to each other or to zero, they are small relative to the norm of the matrix, and
their associated right eigenvectors are almost identical, indicating the nearby presence of a double eigenvalue.
Furthermore, the diagonal and upper triangular elements are very small compared to the norm of the matrix, implying that
a relatively small perturbation removing them yields a Jordan block with a double zero eigenvalue.

Finally there is a fourth flavour of results: those that did not even reach a good approximation to the locally optimal value $\kappa_{0}$.

A final comment on Fig.\ \ref{fig:n2Beta1LimPiOver2}: the \granso\ termination codes are plotted at the bottom of the first panel.
The value 1 means that \granso\ terminated because the limit of 500 iterations was reached, while the value 2 means that it terminated
because it could not find a higher feasible value for the load. Observe that the latter termination always occurred for the runs which approximated
the apparent globally optimal load $\kappa_{0}+\pi$ well (the first flavour) and the runs that obtained loads higher than $\kappa_{0}$
but  terminated without reaching a good approximation to $\kappa_{0}+\pi$ (the third flavour).
Thus, increasing the iteration limit would not have improved any of these values. On the other hand, the runs that provided a
good approximation to the locally maximal value $\kappa_{0}$ (the second flavour) or terminated before reaching that value (the fourth flavour)
sometimes, but not always, terminated by exceeding the maximum iteration limit.

The physical interpretation of the proposed supremum \rf{supremum-kappa-alpha}, \rf{supremum-beta} is that
the mass $M_{2}$ mounted on the free end of the column is zero in the limit $\kappa\to\kappa^{*}$ (assuming that $M_{1}$ is bounded above).
It's interesting to consider what happens if we disallow this case, putting
an upper limit on $\mu_{1}=M_{1}/M_{2}$. Fig.\ \ref{fig:n2Beta1LimArctan100-10} shows the results when we
introduce the constraint $\mu_{1}\leq 100$  (left) or $\mu_{1}\leq 10$ (right) by limiting $\beta_{1}$ to $\arctan(100)$
or $\arctan(10)$ respectively. For $\mu_{1}\leq 100$, we now find an optimal load of $7.6287$, and for $\mu_{1}\leq 10$, we find
the optimal load $7.4666$, which are respectively just  0.1\%  and 2\% lower than the
apparent unconstrained supremum $\kappa_{0}+\pi$.
The biggest difference we observe from comparing Fig.\ \ref{fig:n2Beta1LimArctan100-10} with
Fig.\ \ref{fig:n2Beta1LimPiOver2} is that the eigenvalues of the final computed $\M$ are now dramatically reduced, from more than $10^{16}$
to less than 100 and 15 respectively.
Thus, we obtain an only slightly reduced optimal load while introducing a much more physically reasonable model
with much better numerical properties.

\begin{figure}
 \begin{center}
 \begin{tabular}{cc}
 \includegraphics[scale=0.39,trim={1.5cm 1.5cm 0.5cm 1.5cm},clip]{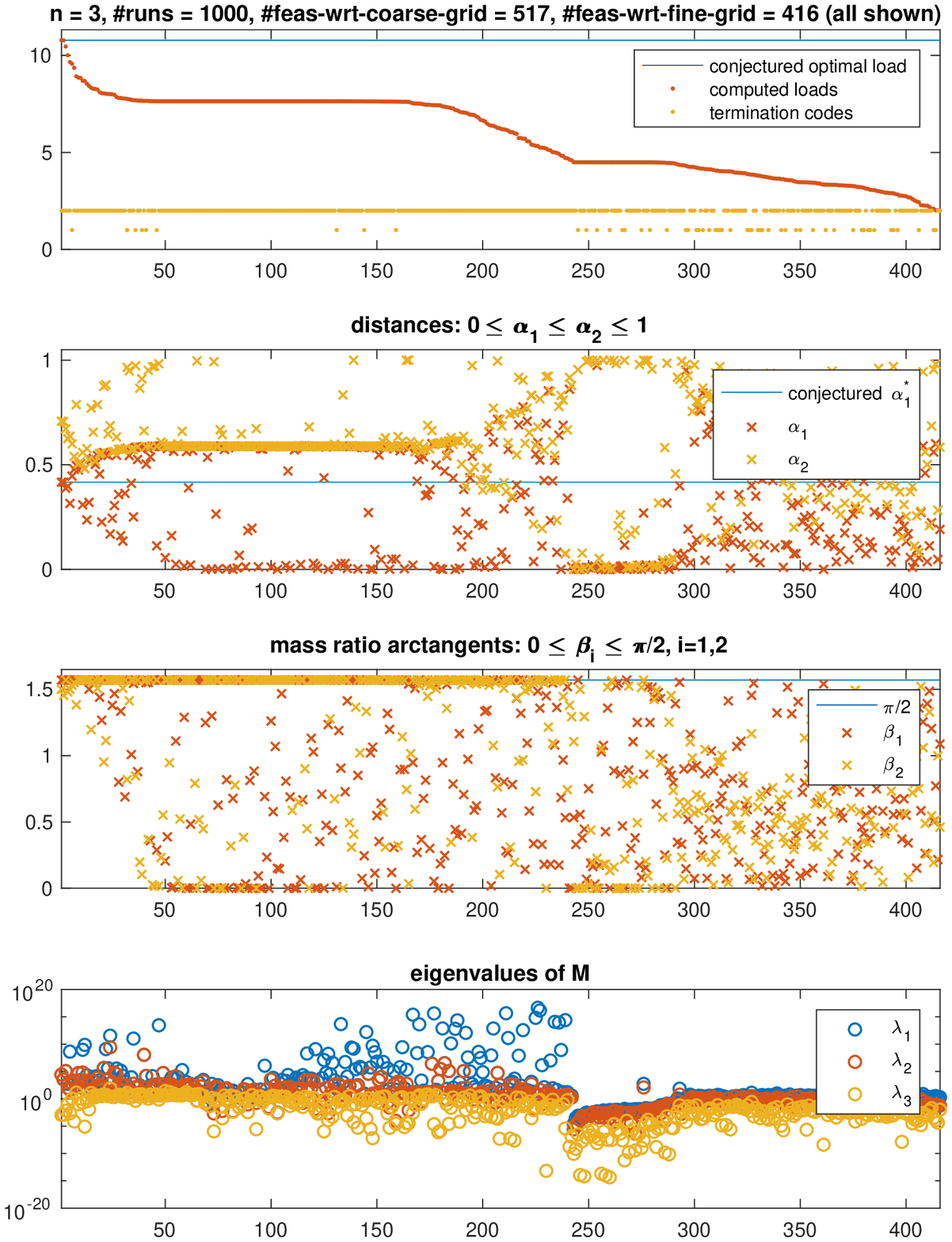} &
 \includegraphics[scale=0.39,trim={1.5cm 1.5cm 0.5cm 1.5cm},clip]{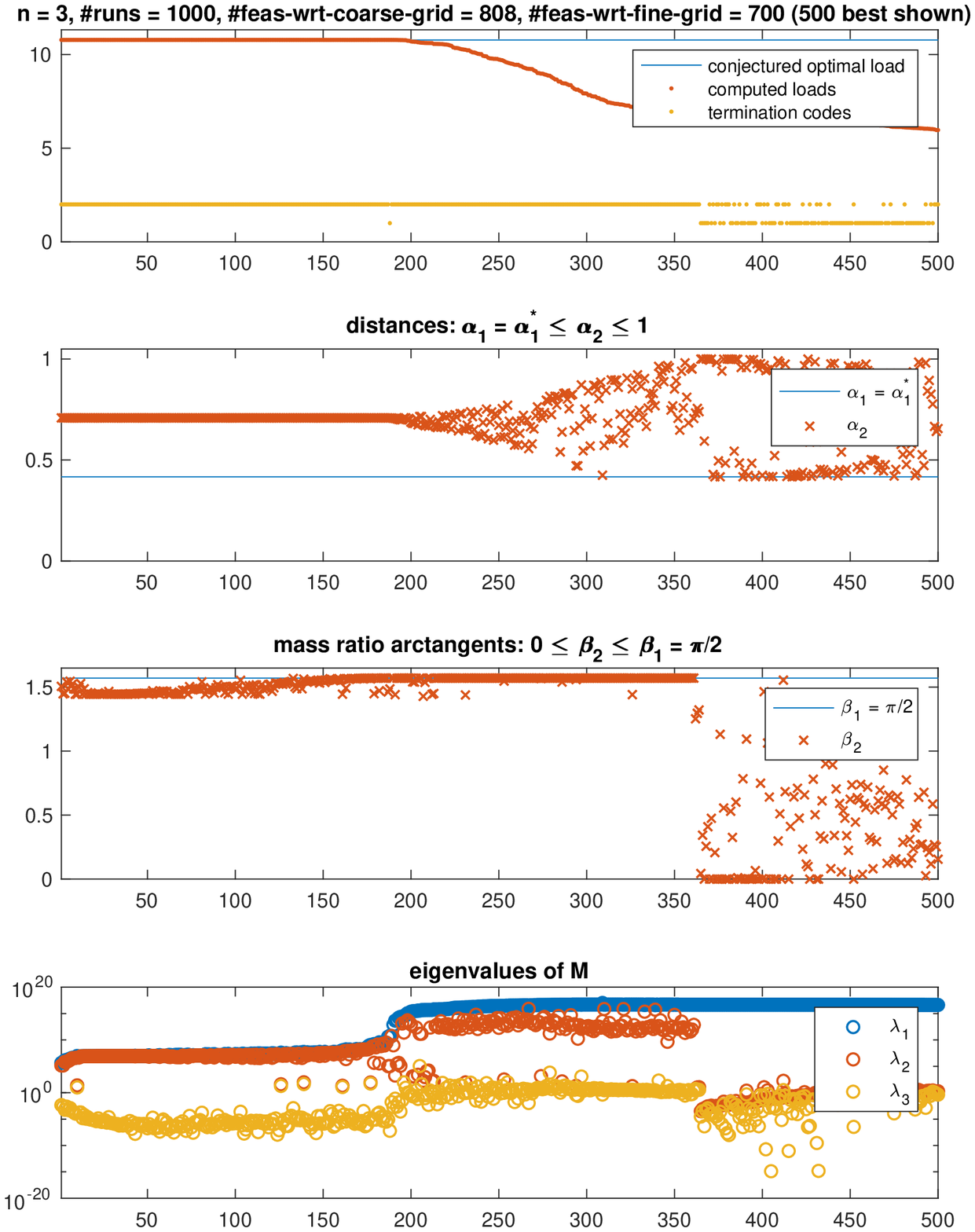}
 \end{tabular}
 \end{center}
 \caption{Solving \eqref{optprob} for $n=3$ with (left) no additional constraints
 and (right) with the constraints $\alpha_{1}=\alpha_{1}^{*}=\kappa_0[\kappa_0+2\pi]^{-1}$, $\beta_{1}=$ {\tt pi/2}.
 See the accompanying text for more details.}
 \label{fig:n3alpha1beta1free-alpha1beta1fixed}
 \end{figure}

  \begin{figure}
 \begin{center}
 \begin{tabular}{cc}
 \includegraphics[scale=0.39,trim={1.5cm 1.5cm 1.3cm 1.5cm},clip]{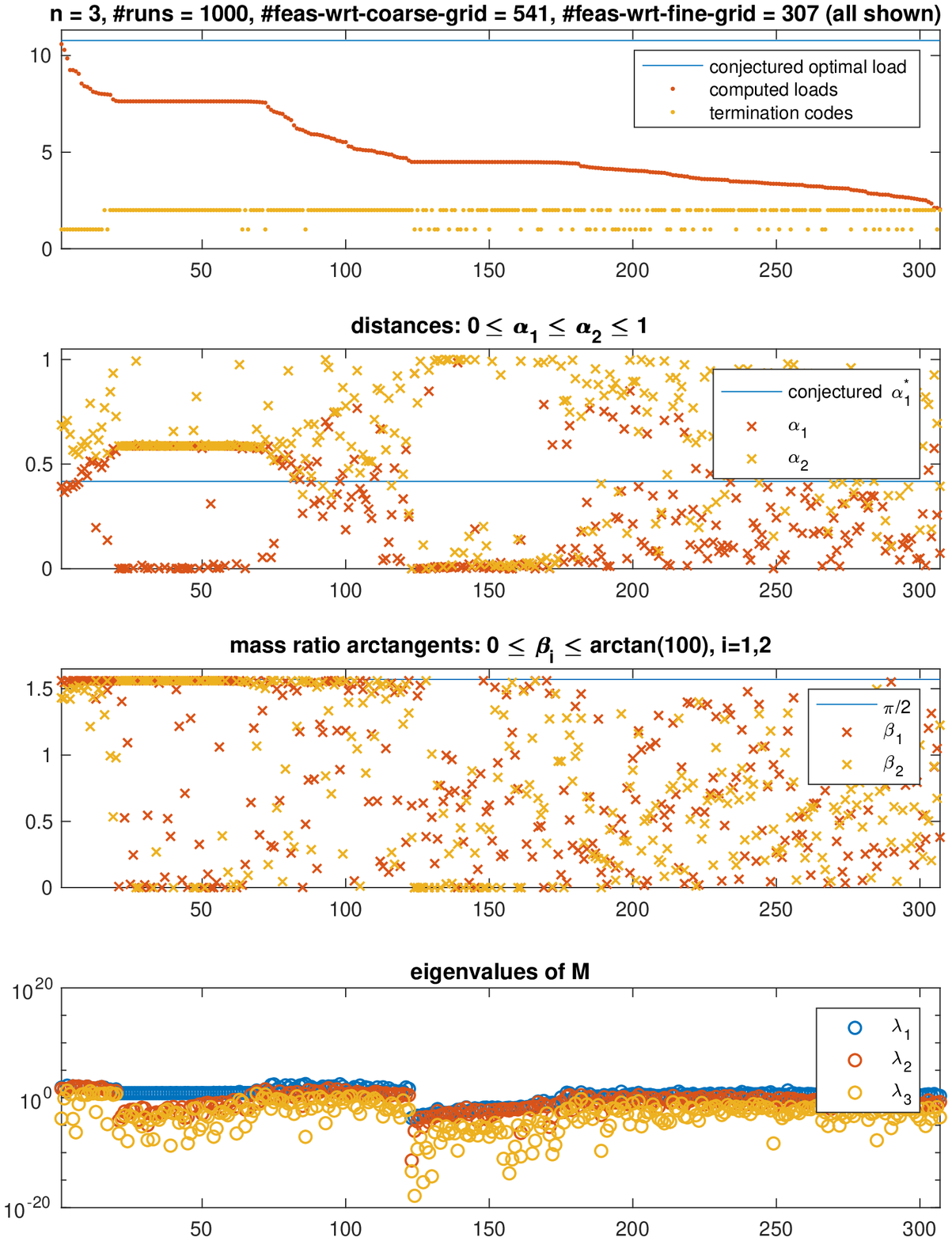} &
 \includegraphics[scale=0.39,trim={1.5cm 1.5cm 1.3cm 1.5cm},clip]{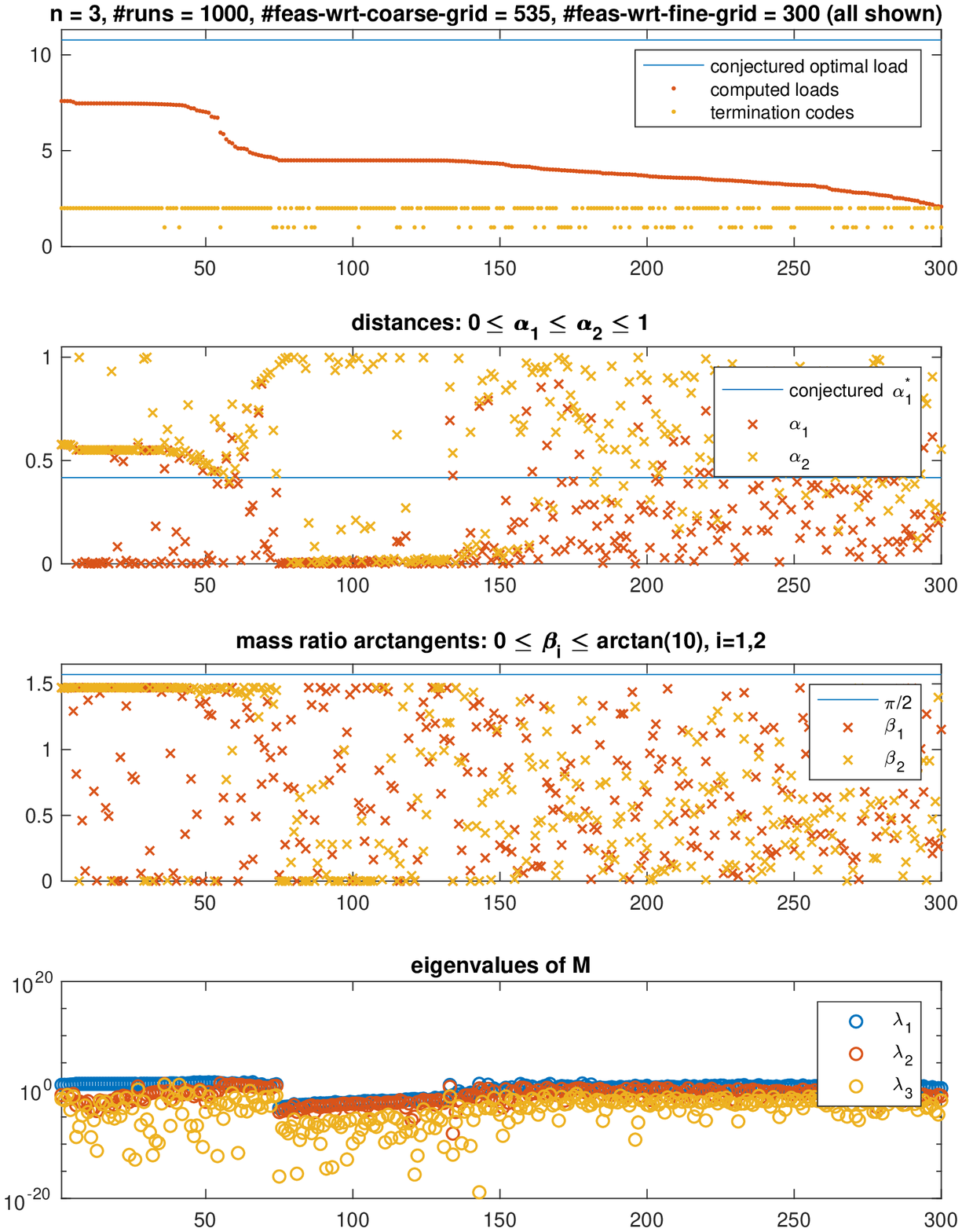}
 \end{tabular}
 \end{center}
 \caption{Solving \eqref{optprob} for $n=3$ with the additional constraints (left) $\beta_{i}\leq \arctan(100)$, $i=1,2$
and (right) $\beta_{i}\leq \arctan(10)$, $i=1,2$.}
 \label{fig:n3BetaLimArctan100-10}
 \end{figure}

\subsection{Results for $n=3$}

The left panel in Fig.\ \ref{fig:n3alpha1beta1free-alpha1beta1fixed} shows the results for solving \eqref{optprob} for $n=3$.
There are five variables: $\alpha_{1},\alpha_{2},\beta_{1},\beta_{2}$ and $\kappa$.
Of the 1000 candidate solutions generated by \granso, 517 satisfied the bound and coarse grid stability constraints imposed by \granso,
and of these, 416 passed the fine grid test. We see immediately that the problem for $n=3$ is significantly harder
than for $n=2$, with not many runs approximating the conjectured optimal value well.
Nonetheless, the two best runs generate $\kappa\approx 1.0776$, which agrees with the conjectured optimal
value $\kappa_{0}+2\pi$ to five digits. These two runs also generate $\alpha_{1}\approx 0.4169$ and $\beta_{1}\approx 1.570796$
which agree with the conjectured optimal values $\alpha_{1}^{*}=\kappa_0[\kappa_0+2\pi]^{-1}$ and $\pi/2$ to 4 and 7 digits,
respectively. The right panel in the same figure shows the results when we fix $\alpha_{1}= \alpha_{1}^{*}$ and $\beta_{1}=$ {\tt pi/2}
(the 16 digit rounded value of $\pi/2$) and
optimize over the remaining three variables $\alpha_{2}$, $\beta_{2}$ and $\kappa$. Then the best two runs generate $\kappa$
agreeing with $\kappa_{0}+2\pi$ to 12 digits, and the best 100 runs agree with this to 10 digits. Together, the results reported in
the left and right panels of Fig.\ \ref{fig:n3alpha1beta1free-alpha1beta1fixed} make a convincing argument that the values shown
in \rf{kapm} and \rf{kapm1} are indeed the supremal value $\kappa^{*}$ and the corresponding limiting value $\alpha_{1}^{*}$ when $n=3$, and that the
corresponding limiting value $\beta_{1}^{*}$ is again $\pi/2$.
Although we do not have a conjectured formula for $\alpha_{2}^{*}$, its computed optimal value is 0.7085. Furthermore, the limiting value
$\beta_{2}^{*}$ is again apparently $\pi/2$, meaning the mass ratio $\mu_{2}=M_{2}/M_{3} \to\infty$ as $\kappa\to\kappa^{*}$,
which indeed must be the case assuming the limiting value of $M_{2}$ is nonzero and $M_{1}$ is bounded above, since then
$\beta_{1}\to\pi/2$ implies that $M_{3}\to 0$.

Fig.\ \ref{fig:n3BetaLimArctan100-10} shows the results when we
introduce the mass ratio constraint $\mu_{i}\leq 100$  (left) or $\mu_{i}\leq 10$ (right) by limiting $\beta_{i}\leq \arctan(100)$, $i=1,2$
or $\beta_{i} \leq \arctan(10)$, $i=1,2$ respectively. For $\mu_{i}\leq 100$, we now find an optimal load of 10.589,
which is only 0.5\% lower than $\kappa_{0}+2\pi$. However, when we constrain $\mu_{i}\leq 10$, the best optimal load found is only 7.59,
which is a 30\% reduction from $\kappa_{0}+2\pi$.
When we repeat these runs with 10,000 starting points instead of 1000, these numbers are only slightly improved.

 \begin{figure}
 \begin{center}
 \begin{tabular}{cc}
 \includegraphics[scale=0.38,trim={1.5cm 1.5cm 1.3cm 1.5cm},clip]{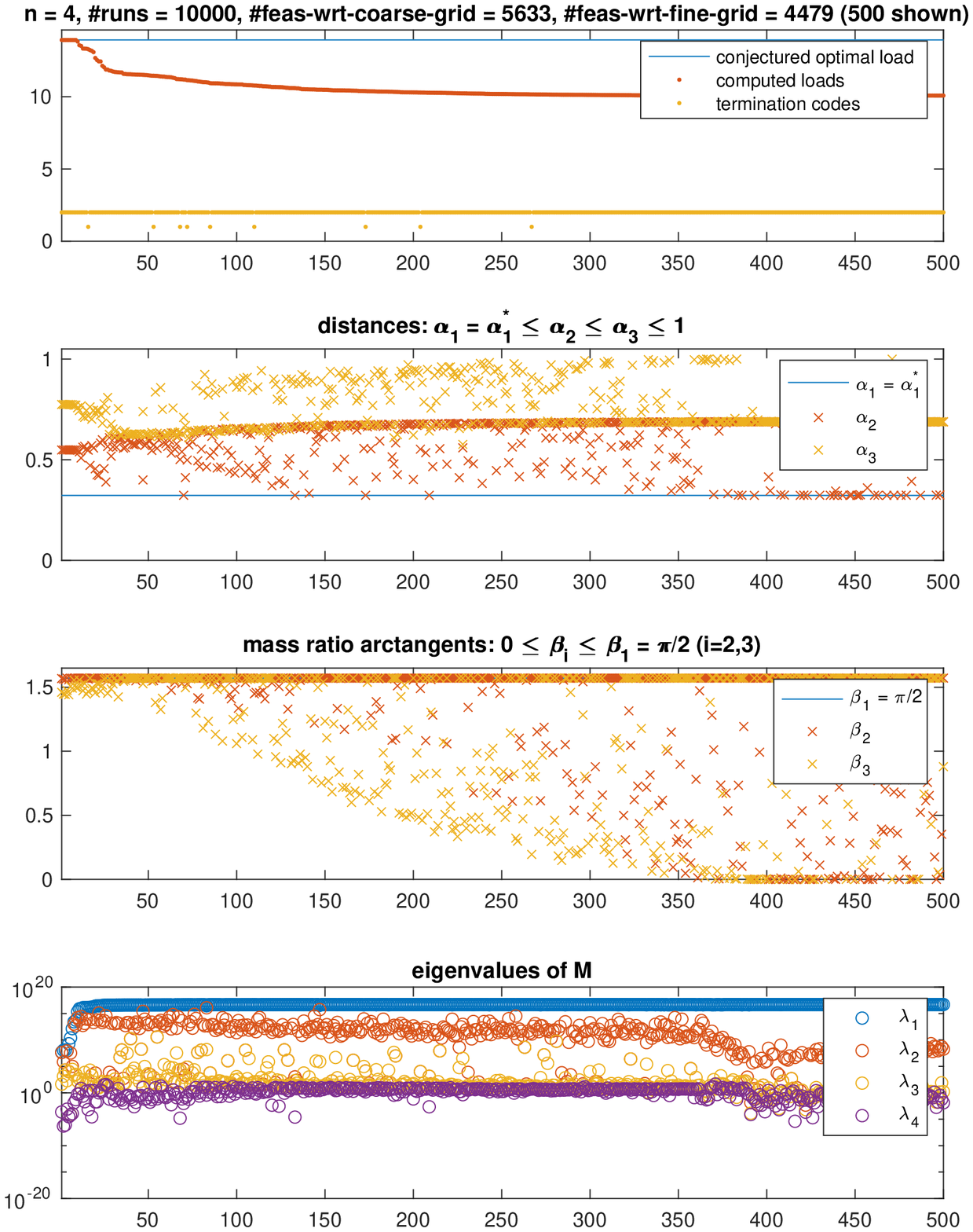} &
 \includegraphics[scale=0.38,trim={1.5cm 1.5cm 1.3cm 1.5cm},clip]{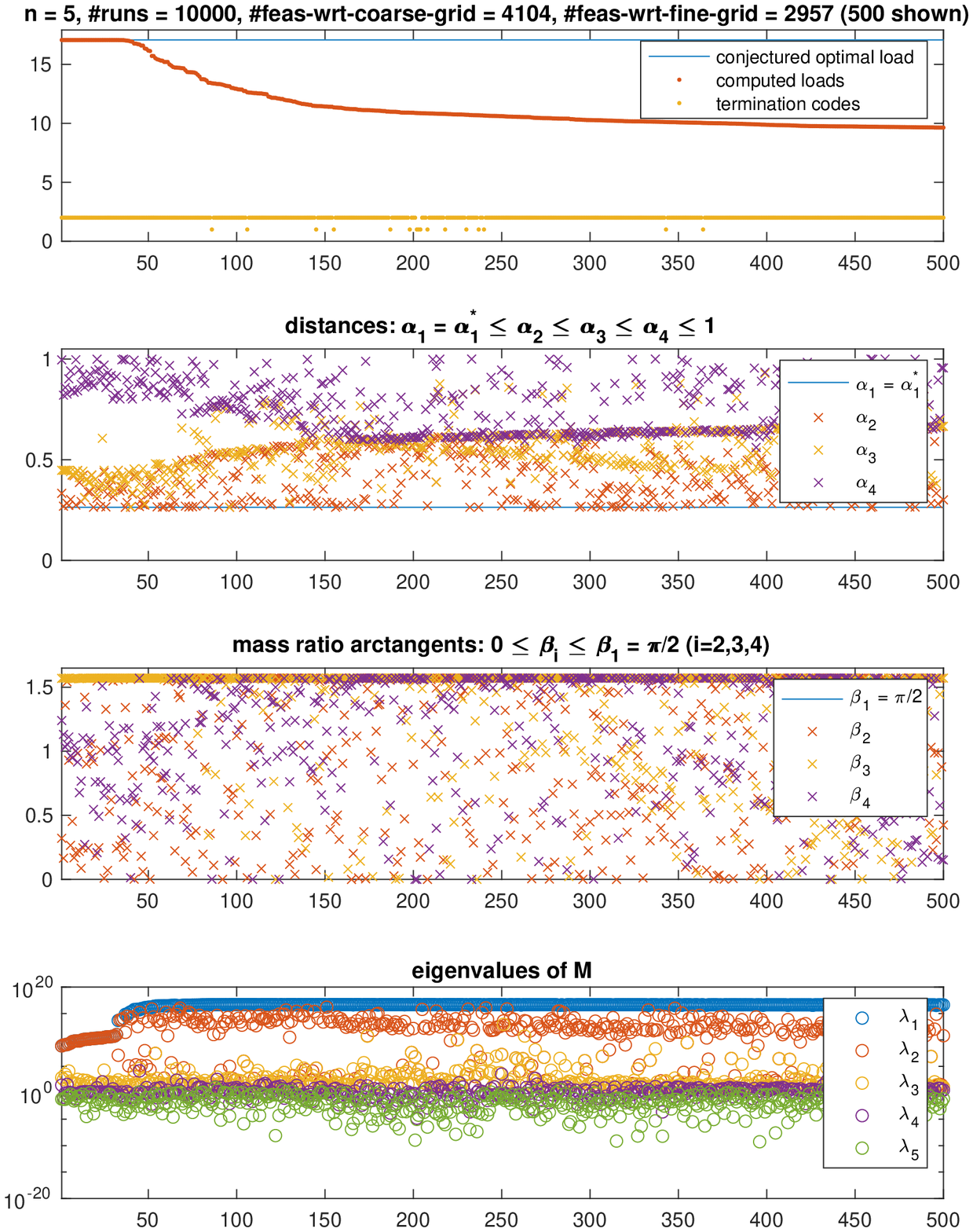}
 \end{tabular}
 \end{center}
  \caption{(Left) solving \eqref{optprob} for $n=4$ with the constraints $\alpha_{1}=\alpha_{1}^{*}=\kappa_0[\kappa_0+3\pi]^{-1}$ ,
  $\beta_{1}=$ {\tt pi/2} and (right)  solving \eqref{optprob} for $n=5$ with the constraints
  $\alpha_{1}=\alpha_{1}^{*}=\kappa_0[\kappa_0+4\pi]^{-1}$, $\beta_{1}=$ {\tt pi/2}.}
 \label{fig:n4-5}
 \end{figure}

 \subsection{Results for $n=4$ and $n=5$}

 The optimization problem is so much harder for $n=4$ and $n=5$ that we needed 10,000 starting points to get good results,
 even when we set $\alpha_{1}$ to its conjectured optimal value in \rf{kapm1}
 and $\beta_{1}$ to {\tt pi/2}, optimizing over the remaining 5 and 7 variables, respectively.
 The results are shown in the left and right panels of Fig.\ \ref{fig:n4-5}.
 For $n=4$, the best 5 results agree with our conjectured optimal load $\kappa_{0} + 3\pi$ to 8 digits,
 while for $n=5$, the best 25 results agree with $\kappa_{0}+4\pi$ to 7 digits.
These results strongly support our conjecture regarding
 the supremal load $\kappa$  for $n$ masses given in \rf{kapm}.

 \section{Concluding Remarks}
 We believe we have made a convincing case that the supremal load for the
 strongest stable massless column with a follower load and $n$ relocatable masses is, in the dimensionless model defined
 in Section \ref{sec:weightless}, $\kappa_{0}+(n-1)\pi$, where $\kappa_{0}$ is the smallest positive
 root of $\tan(\kappa)=\kappa$.
 This conjecture has not previously appeared in the literature as far as we know, except in the case $n=1$ where it
 has been known to be true for decades  \cite{B1963}. We have given a detailed analytical derivation of this result for $n=2$, and
 presented extensive computational results that support it for $n=2,3,4,5$, using numerical nonsmooth optimization.

 With this model problem effectively solved, we believe it would be interesting to apply our nonsmooth optimization techniques
 to more realistic columns with follower loads \cite{GZ1988}, such as the Beck, Pfl\"uger and Leipholz columns with a single free end as well as to free-free beams both with distributed and concentrated masses to get new insights about the nature of the optimal solution to these long-standing optimization problems. We believe it is also important to consider extending
 traditional stability constraints to more robust stability constraints based on pseudospectra \cite{TE2006},
 a topic that is beyond the scope of this paper.

\smallskip
\noindent
{\bf Acknowledgements.} The authors thank Tim Mitchell, the author of \granso, for many helpful discussions and suggestions
regarding the formulation of the stability constraint. They also thank the London Mathematical Society for
supporting the second author's visit to Northumbria through the Scheme 4 Research in Pairs grant No 41820.
The second author was supported in part by the U.S.\ National Science Foundation Grant DMS-2012250.

\begin{appendix}

\section{Pfl\"uger's column}\label{PflugerApp}

To model the Pfl\"uger column we consider an elastic beam of length $l$, with Young's modulus $E$
and mass per unit length $m$, clamped at one end and loaded by a tangential follower force $P$ at the other end, where a point mass $M$ is mounted. The moment of inertia of a cross-section of the column is denoted by $I$.
Small lateral vibrations of the Pfl\"uger column near the undeformed equilibrium are described by the
linear partial differential equation \cite{P1955,TKMB2016}
\begineq{pce}
EI\frac{\partial^4 y}{\partial s^4}+P\frac{\partial^2 y}{\partial s^2}+m\frac{\partial^2 y}{\partial t^2}=0
\end{equation}
where $y(s,t)$, is the amplitude of the vibrations and $s \in [0, l]$ is a coordinate along the column.
At the clamped end $(s = 0)$ equation \rf{pce} satisfies the boundary conditions
\begineq{pbc1}
y=0, \quad \frac{\partial y}{\partial s}=0,\quad s=0,
\end{equation}
while at the loaded end $(s = l)$, the boundary conditions are
\begineq{pbc2}
EI\frac{\partial^2 y}{\partial s^2}=0,\quad EI\frac{\partial^3 y}{\partial s^3}=M\frac{\partial^2 y}{\partial t^2}, \quad s=l.
\end{equation}

    \begin{figure}
    \begin{center}
    \includegraphics[angle=0, width=0.99\textwidth]{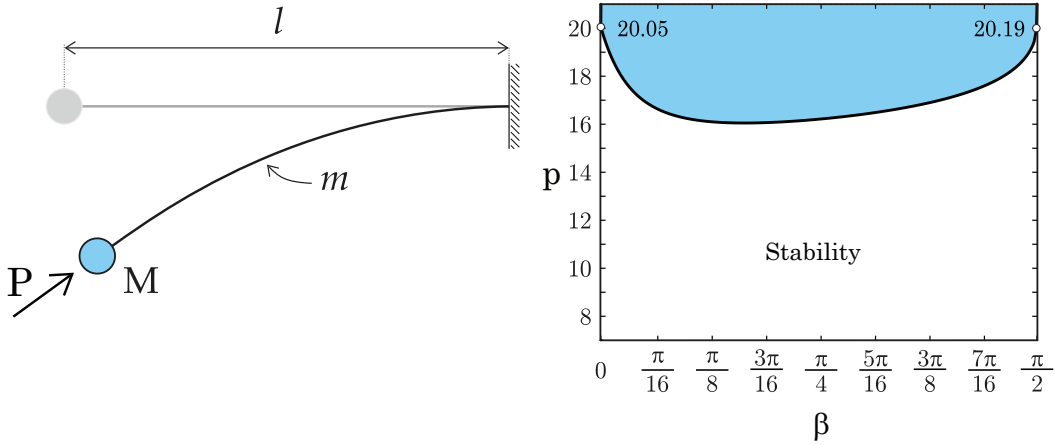}
    \end{center}
    \caption{The Pfl\"uger column and its stability diagram. The ratio of the end mass to the mass of the column, $\mu=M/(ml)$, is parameterized by $\mu=\tan\beta$. The Beck column corresponds to the vanishing end mass ($M=0$, so $\beta=0$) and the massless Pfl\"uger column (or Dzhanelidze's column \cite{B1963}) to the vanishing mass of the rod ($m=0$, so $\beta=\pi/2$).
The vertical axis of the stability diagram shows the dimensionless load $p=\frac{Pl^2}{EI}$, where $E$ is Young's modulus, $I$ is the moment of inertia of a cross-section of the column and $l$ is the the length of the column.}
    \label{fig1}
    \end{figure}

Introducing the dimensionless quantities
\begineq{dlq}
\xi=\frac{s}{l},\quad \tau=\frac{t}{l^2}\sqrt{\frac{EI}{m}},\quad p=\frac{P l^2}{EI},\quad \mu=\frac{M}{ml},
\end{equation}
and separating the time variable through $y(\xi, \tau) = lf(\xi)\exp(\lambda \tau)$, we obtain the dimensionless boundary eigenvalue problem
\begineq{dlbp}
\partial_{\xi}^4 f +p \partial_{\xi}^2 f +\lambda^2 f=0,
\end{equation}
\ba{dlbc}
&\partial_{\xi}^2 f(1)=0,\quad \partial_{\xi}^3 f(1)=\mu \lambda^2 f(1),&\nn \\
&f(0)=0,\quad \partial_{\xi}f(0)=0&
\ea
defined on the interval $\xi \in [0, 1]$.
A solution to the equation \rf{dlbp} with boundary conditions \rf{dlbc} is \cite{P1955,TKMB2016}
\begineq{sbp}
f(\xi)=A(\cosh(g_2 \xi)-\cos(g_1 \xi))+B(g_1\sinh(g_2 \xi)-g_2\sin(g_1 \xi))
\end{equation}
with
$$
        g_{1,2}=\sqrt{\frac{\sqrt{p^2-4\lambda^2}\pm p}{2}},
$$
where the subscripts $1$ and $2$ correspond to the signs $+$ and $-$, respectively.
Imposing the boundary conditions \rf{dlbc} on the solution \rf{sbp} yields the characteristic equation $\Delta(\lambda) = 0$ for the
determination of the eigenvalues $\lambda$, where
$$
\Delta(\lambda)=\Delta_1-\Delta_2\mu \lambda^2
$$
and
\ba{}
\Delta_1&=&g_1g_2(g_1^4+g_2^4+2g_1^2g_2^2\cosh g_2 \cos g_1 + g_1 g_2 (g_1^2-g_2^2)\sinh g_2 \sin g_1) \nn\\
\Delta_2&=&(g_1^2+g_2^2)(g_1 \sinh g_2 \cos g_1 - g_2 \cosh g_2 \sin g_1).
\ea

Parameterizing the mass ratio
in \rf{dlq} by $\mu = \tan \beta$ with $\beta \in [0, \pi/2]$ enables the exploration of all possible
ratios  $\mu=M/(ml)$ of the end mass to the mass of the column from zero ($\beta = 0$) to infinity ($\beta = \pi/2$).
The former case, without the end mass, corresponds to the Beck column, whereas the latter corresponds to a massless rod
with an end mass, which is known as the Dzhanelidze column \cite{B1963}.

It is well-established that the uniform Beck column is stable against flutter if the dimensionless follower force, $p$, is such that $0\le p \lesssim 20.05$,  \cite{B1952,CM1979,B1963,K2013}. In contrast, the Dzhanelidze column becomes unstable at $p \approx 20.19$, which is the smallest
positive root of the equation \cite{B1963}
\begineq{dcp}
\tan \sqrt{p} = \sqrt{p}.
\end{equation}
These values, representing two extreme situations, are connected by a marginal stability
curve in the $(\beta, p)$-plane  \cite{B1963,P1955,TKMB2016,O1972,SKK1976,CK1992}; see the right panel of Fig.~\ref{fig1}.

For every fixed value $\beta \in [0, \pi/2)$, the Pfl\"uger column loses stability via flutter when an increase in $p$ causes the
imaginary eigenvalues of two different modes to approach each other and merge into a double imaginary eigenvalue with
a Jordan block (i.e., with algebraic multiplicity two and geometric multiplicity one).
When $p$ crosses the threshold, the double eigenvalue splits into two complex eigenvalues, one with
positive real part, which determines a flutter-unstable mode.

At $\beta = \pi/2$ the stability boundary of the Pfl\"uger column has a vertical tangent and the type of instability
changes from flutter to divergence, i.e., non-oscillatory growth of a mode corresponding to a positive real eigenvalue, for $p \gtrsim 20.19$; see \cite{B1963,O1972,SKK1976}.

\end{appendix}

\end{document}